\documentclass[a4paper,11pt,preprintnumbers]{article}

\usepackage{physics}
\usepackage{makecell}
\usepackage{amssymb,mathrsfs,amsmath}
\usepackage{a4wide}
\usepackage{color}
\usepackage{slashed,soul}
\usepackage{graphicx}
\usepackage{amsfonts}
\usepackage{lscape}

\usepackage{jheppub}

\hypersetup{%
  colorlinks=true, linktocpage=true, pdfstartpage=1, pdfstartview=FitV,
  breaklinks=true, pageanchor=true,
  pdfpagemode=UseNone,
  plainpages=false, bookmarksnumbered, bookmarksopen=true, bookmarksopenlevel=1,
  hypertexnames=true, pdfhighlight=/O,
  urlcolor=RoyalBlue, linkcolor=RoyalBlue, citecolor=RoyalBlue,
  pdftitle={Hunting for ALP production in meson and gauge boson LFV decays},
  pdfauthor={Marco Ardu, Lorenzo Calibbi, Marco Fedele, Federico Mescia},
  pdfsubject={},
  pdfkeywords={},
  pdfcreator={pdfLaTeX},
  pdfproducer={LaTeX with hyperref}
}

\usepackage{colortbl,colordvi,color}
\usepackage[table,dvipsnames]{xcolor}
\usepackage{amsthm}
\usepackage{booktabs}
\usepackage{array}
\usepackage{rotating}
\usepackage[numbers, sort&compress]{natbib}
\usepackage[normalem]{ulem}
\usepackage{multirow}
\usepackage{float}
\usepackage[utf8]{inputenc}
\usepackage[T1]{fontenc}
\usepackage{appendix}
\usepackage{epsfig}
\usepackage{orcidlink}
\usepackage{comment}
\usepackage{adjustbox}
\usepackage{tcolorbox}
\usepackage[section]{placeins}

\newcommand{\beq}{\begin{equation}} 
\newcommand{\eeq}{\end{equation}} 
\newcommand{\ba}{\begin{array}}  
\newcommand{\ea}{\end{array}} 
\newcommand{\bea}{\begin{eqnarray}}  
\newcommand{\eea}{\end{eqnarray} }  
\newcommand{\bal}{\begin{align}}
\newcommand{\eal}{\end{align}}   
\newcommand{\bi}{\begin{itemize}}  
\newcommand{\ei}{\end{itemize}}  
\newcommand{\ben}{\begin{enumerate}}  
\newcommand{\een}{\end{enumerate}}  
\newcommand{\bc}{\begin{center}}
\newcommand{\ec}{\end{center}} 
\newcommand{\bt}{\begin{table}}
\newcommand{\et}{\end{table}}  
\newcommand{\btb}{\begin{tabular}}
\newcommand{\etb}{\end{tabular}}

\renewcommand{\O}{\mathcal{O}}

\newcommand{\fref}[1]{Figure~\ref{#1}}

\newcommand{\GeV}{\,\mathrm{GeV}} 
\newcommand{\TeV}{\,\mathrm{TeV}}

\definecolor{orange}{rgb}{1,0.5,0}

\allowdisplaybreaks

\let\OLDthebibliography\thebibliography
\renewcommand\thebibliography[1]{
  \OLDthebibliography{#1}
  \setlength{\parskip}{0pt}
  \setlength{\itemsep}{0pt plus 0.3ex}
}

\begin{document}
\vspace{1cm}

\begin{titlepage}

\begin{flushright}
MITP-26-018
 \end{flushright}
\vspace{0.2truecm}

\begin{center}
\renewcommand{\baselinestretch}{1.8}\normalsize
\boldmath
{\LARGE\textbf{
ALP production 
in Lepton Flavour Violating\\ meson, tau and gauge boson decays
}}
\unboldmath
\end{center}

\vspace{0.4truecm}

\renewcommand*{\thefootnote}{\fnsymbol{footnote}}

\begin{center}

{
Marco Ardu\,$^1$\footnote{\href{mailto:marco.ardu@ific.uv.es}{marco.ardu@ific.uv.es}}, Lorenzo Calibbi\,$^2$\footnote{\href{mailto:calibbi@nankai.edu.cn}{calibbi@nankai.edu.cn}}, 
Marco Fedele\,$^3$\footnote{\href{mailto:mfedele@uni-mainz.de}{mfedele@uni-mainz.de}}, Federico Mescia\,$^4$\footnote{\href{mailto:federico.mescia@lnf.infn.it}{federico.mescia@lnf.infn.it}}
}
\vspace{0.7truecm}

{\footnotesize
$^1$ Instituto de F\'{\i}sica Corpuscular, Universidad de Valencia and CSIC\\ 
 Edificio Institutos Investigaci\'on, Catedr\'atico Jos\'e Beltr\'an 2, E-46980 Spain
 \\[1mm]
$^2$ School of Physics, Nankai University, Tianjin 300071, China
 \\[1mm]
$^3$ PRISMA$^{++}$ Cluster of Excellence \& Mainz Institute for Theoretical Physics \\
Johannes Gutenberg University, D-55099 Mainz, Germany \\[1mm]
$^4$ Istituto Nazionale di Fisica Nucleare, Laboratori Nazionali di Frascati, C.P.~13, I-00044 Frascati, Italy
 \\
}
\vspace*{2mm}
\end{center}

\renewcommand*{\thefootnote}{\arabic{footnote}}
\setcounter{footnote}{0}

\begin{abstract}

In this paper we study axion-like particles (ALPs) with lepton-flavour-violating (LFV) couplings in the mass regime above the muon threshold, $m_a>m_\mu$, where the strong bound from the exotic muon decay \mbox{$\mu\to ea$} no longer apply and the decay channel \mbox{$a\to e\mu$} becomes kinematically accessible. In this region, the ALP typically decays promptly, motivating new search strategies based on its production in decays involving virtual muons. We analyse charged-meson and $W$ decays, neutral-current processes such as $Z$ and quarkonium decays, and, when couplings to the third generation are present, LFV $\tau$ decays. 
The subsequent decay $a\to e\mu$ leads to striking LFV signatures with negligible Standard Model backgrounds. Combining these production modes with current low-energy constraints, we assess the sensitivity of future high-energy $e^+e^-$ colliders, flavour factories such as Belle~II and STCF, fixed-target experiments such as NA62, and proton beam-dump facilities such as SHiP.
Overall, our results identify LFV ALP production in meson, gauge-boson, quarkonium and $\tau$ decays (with displaced vertices) as a promising and largely unexplored avenue to test ALP interactions with charged leptons above the muon mass threshold.
\end{abstract}

\end{titlepage}

\tableofcontents

\section{Introduction}

Light feebly-coupled pseudo Nambu-Goldstone bosons (PNGBs), such as axions and axion-like particles~(ALPs), are a natural consequence of a large class of new physics~(NP) models entailing the spontaneous breaking of continuous global symmetries.
Strongly motivated examples include models addressing some open questions of the Standard Model~(SM), such as the strong CP problem~\cite{Peccei:1977hh,Weinberg:1977ma,Wilczek:1977pj}, the origin of neutrino masses~\cite{Chikashige:1980ui,Gelmini:1980re,Georgi:1981pg} or the underlying dynamics behind the hierarchical structure of fermion masses and mixing~\cite{Davidson:1981zd,Wilczek:1982rv,Reiss:1982sq,Davidson:1983fy,Chang:1987hz,Berezhiani:1990jj,Berezhiani:1990wn,Ema:2016ops,Calibbi:2016hwq, Berezhiani:1983hm, Berezhiani:1985in, Berezhiani:1989fs}. 
Moreover, such light particles (axions, majorons, familons etc.) can be themselves excellent candidates for dark matter~\cite{Preskill:1982cy,Abbott:1982af,Dine:1982ah,Arias:2012az}\,---\,if so light and weakly-coupled to SM fields to be stable on cosmological time scales\,---\,or can otherwise act as portals to a light dark sector~\cite{Nomura:2008ru,Essig:2013lka,Bharucha:2022lty,Ghosh:2023tyz,Fitzpatrick:2023xks,Armando:2023zwz,DEramo:2025xef,Calibbi:2025rpx,Fiorentino:2026dea}, hence reinforcing their physics case. These appealing features explain why this class of light NP particles has recently become the object of renewed interest and a large number of experimental proposals and possible searches at existing and future experiments (including high-energy colliders and high-intensity frontier facilities) have been put forward~\cite{deBlas:2025gyz,Ai:2025cpj,FCC:2018bvk,Dainese:2019rgk,FCC:2025lpp,CEPCStudyGroup:2018ghi,Alimena:2025kjv}. 

In the present work, we focus on ALPs that couple to SM leptons, for instance arising from the spontaneous breaking of the lepton number or (leptonic) flavour symmetries. Within a large class of models, such ALP-lepton interactions are generically flavour-violating, either due to loop effects~\cite{Garcia-Cely:2017oco,Heeck:2019guh} or because of the flavour non-universality of the underlying broken symmetry~\cite{Calibbi:2020jvd} or because of their non-Abelian nature~\cite{Calibbi:2025rxn}. This opens up the exciting possibility of discovering ALPs (and light NP in general) at the numerous running and upcoming experiments aimed at probing exotic lepton flavour-violating (LFV) processes~\cite{Bernstein:2013hba,Calibbi:2017uvl,Ardu:2022sbt,Davidson:2022jai}.

Muon and tau decays into a light ALP $a$, such as $\mu\to e\,a$ and $\tau\to\ell\,a$ (where $\ell = e,\mu$), already set stringent constraints on LFV ALP couplings, and even higher sensitivities are expected to be reached in running and upcoming experiments, 
both in scenarios where the ALP decays visibly inside the detector~\cite{Echenard:2014lma, Heeck:2017xmg, Hostert:2023gpk, Knapen:2023iwg, Fox:2024kda, Knapen:2024fvh} and where it is long‑lived and therefore escapes detection~\cite{GarciaiTormo:2011cyr, Uesaka:2020okd, Panci:2022wlc, Jho:2022snj, Xing:2022rob, Hill:2023dym, Knapen:2023zgi, Bigaran:2025uzn, Calibbi:2020jvd}.
In contrast, if the ALP is heavy enough for these decays to be kinematically forbidden, the current bounds on LFV ALP interactions become significantly weaker. In particular, in this regime, the main direct limits on $e\mu$ ALP interactions stem from searches for muonium to antimuonium conversion and measurements of the muon and electron $g-2$. However, the sensitivity of these processes is currently modest\,---\,see, for instance, Refs.~\cite{Endo:2020mev, Cornella:2019uxs,Bauer:2021mvw,Calibbi:2024rcm}\,---\,as they are suppressed by a double insertion of the small ALP coupling, which is proportional to the muon mass, and inversely proportional to the ALP decay constant $f_a$ that is of the order of the high-energy scale where the underlying global symmetry is spontaneously broken. Proposals to target the kinematic region \mbox{$m_a > m_\mu$} of LFV ALPs have been put forward for Belle~II~\cite{Endo:2020mev}, the proposed $e^-\mu^+$ collider $\mu$TRISTAN~\cite{Calibbi:2024rcm}, and $e^-\gamma$ collisions at running/future high-intensity $e^+e^-$ facilities~\cite{Zhang:2026dui}. However, the estimated sensitivity of these proposed searches does not typically exceed $f_a \approx 10\,\mathrm{TeV}$.\footnote{ALP emission through LFV processes in supernova explosions can provide additional sensitivity to $e\mu$ ALP interactions~\cite{Calibbi:2020jvd}, including within the $m_a\gtrsim m_\mu$ regime~\cite{Li:2025beu,Huang:2025rmy,Huang:2025xvo}.}
In addition, if the flavour-conserving ALP coupling to muons is also present with a strength comparable to the LFV one, a muon-ALP loop can give rise to the LFV decay $\mu\to e\gamma$~\cite{Cornella:2019uxs,Bauer:2021mvw}. As a consequence, the above searches would face strong indirect bounds from the experimental limits on $\mu\to e\gamma$\,---\,in particular the recent result from MEG~II~\cite{MEGII:2025gzr}\,---\,as shown, for instance, in Ref.~\cite{Calibbi:2024rcm}. 

The purpose of the present study is to propose and assess the sensitivities of new possible experimental strategies to test this poorly constrained regime, i.e., $m_a > m_\mu$, with a particular focus on the ALP LFV interactions with muons and electrons. To this aim, we consider experimental setups where large numbers of muons can be produced in decays of mesons and gauge bosons, so as to obtain an enhanced probability that an on-shell ALP is emitted by (virtual) muons converting to electrons: $K\to \mu^* \,(\to e a)\nu_\mu $, $D_s\to \mu^* \,(\to e a)\nu_\mu$, $Z\to \mu \mu^*\,(\to e a)$ etc. This strategy has been in part explored in the case of flavour-conserving couplings in Ref.~\cite{Altmannshofer:2022ckw}. However, the combination with LFV signatures has not yet been considered. In particular, as we will show, the subsequent LFV ALP decay $a \to e\mu$ can lead to striking background-free signatures{, such as $Z \to e^\pm \mu^\mp e^\pm \mu^\mp$, which extends to the case of on-shell particle decays the analogous Belle~II $e^+e^- \to e^\pm e^\pm \mu^\mp \mu^\mp$ search proposed in Ref.~\cite{Endo:2020mev}}. Combined with the large number of mesons and electroweak gauge bosons expected to be produced at existing and future high-luminosity facilities\,---\,including NA62~\cite{NA62:2017rwk}, SHiP~\cite{SHiP:2025ows}, the STCF~\cite{Achasov:2023gey}, and the CEPC/FCC-ee~\cite{CEPCStudyGroup:2023quu,FCC:2025lpp}\,---\,this approach offers a promising opportunity to explore an otherwise unconstrained region of parameter space.

If the ALP has LFV
couplings to the $\tau$ lepton, an additional production channel can be considered through the LFV $\tau$ decay to an on-shell ALP, $\tau \to \ell a\, (\to \mu e)$. The observation of this decay chain in flavour-factory experiments such as Belle~II~\cite{Belle-II:2018jsg} and the measurement of the ALP lifetime and branching ratios would then provide indirect sensitivity to $e\mu$ ALP interactions, enabling an indirect test of $e\mu$ lepton flavour violation with $\tau$ decays. A similar search for $\tau \to \ell a$ followed by the displaced flavour-conserving ALP decays $a\to \mu \mu,\,ee$ at Belle~II has been proposed in Ref.~\cite{Cheung:2021mol}.

The rest of the paper is organised as follows. In Section~\ref{sec:setup}, we introduce the setup of the effective LFV ALP framework and establish the notation used throughout this work. Section~\ref{sec:production} presents the core theoretical results: we compute the production rates of axion-like particles in a variety of LFV processes, including meson decays, gauge-boson decays, and $\tau$-lepton decays. In Section~\ref{sec:pheno} we turn to the phenomenology, discussing current low-energy constraints, the prospects at future $e^+e^-$ colliders, and the sensitivity of proton fixed-target and beam-dump experiments, before combining all results into a unified picture. We summarise our conclusions in Section~\ref{conclu}. Additional technical details on model building and simulation procedures are provided in Appendices~\ref{app:weak} and~\ref{app:beamdump}, respectively.

\section{Setup and notation}
\label{sec:setup}

Given the PNGB nature of the ALP, in the low-energy effective field theory, its interactions are governed by an approximate shift symmetry, $ a \to a + \text{const} $, which ensures that only derivative couplings to matter fields appear at leading order. This shift symmetry can be explicitly broken, giving rise to an ALP mass term, and may also be violated by anomaly terms if the underlying global symmetry is anomalous. We here focus on the interactions of the ALP with leptons, for which the relevant Lagrangian terms are 
\begin{equation}
	\mathcal{L}_{a\ell\ell} =
	\sum_i \frac{\partial_\mu a}{2 f_a} \, 
	\bar \ell_i \, C_{ii}^A \, \gamma^\mu \gamma^5 \ell_i
	+ 
	\sum_{i\neq j} \frac{\partial_\mu a}{2 f_a} \, 
	\bar\ell_i \gamma^\mu\left( C_{ij}^V + C_{ij}^A \gamma^5 \right) \ell_j \,,\label{eq:LagrangianCLepton}
\end{equation}
where $\ell_i, \ell_j$ are the charged leptons, $C^{V,A}_{ij}$ are Hermitian matrices of dimensionless coefficients, and the ALP decay constant $f_a$ is an energy scale related to the vacuum expectation value (VEV) of the scalar field that breaks the underlying symmetry. In the following, in order not to commit to a specific model, we will treat these quantities as free parameters.
Notice that the flavour-diagonal vector couplings are set to zero, as the corresponding vector currents are conserved and can be eliminated through field redefinitions. Although our discussion focuses on ALP couplings to charged leptons, it is important to note that in the $SU(2)_L$-symmetric limit the ALP couplings to neutrinos,
\begin{equation}
	\mathcal{L}_{a\nu\nu} =
	\frac{\partial_\mu a}{2 f_a} \, C_{ij}^{\nu} \, 
	\bar \nu_i \gamma^\mu P_L \nu_j \,,\label{eq:ALPneutrinos}
\end{equation}
must be related to the corresponding couplings to the left-handed (LH) charged leptons, namely $C^V_{ij}-C^A_{ij}\overset{SU(2)_L}{=}C^{\nu}_{ij}$. Departures from the $SU(2)_L$-invariant condition can have interesting phenomenological consequences, see Ref.~\cite{Altmannshofer:2022ckw}. Indeed, although it is well known that\,---\,up to anomaly terms~\cite{Eberhart:2025lyu}\,---\,the derivative Lagrangian of Eq.~(\ref{eq:LagrangianCLepton}) is equivalent to the scalar interactions
\begin{equation}
	\mathcal{L}_{a\ell\ell} = i \frac{a}{2 f_a} \bar \ell_i 
	\left[
	(m_{\ell_i} - m_{\ell_j}) C_{ij}^{V}
	+ (m_{\ell_i} + m_{\ell_j}) C_{ij}^{A} \gamma^5
	\right] \ell_j\,,
\end{equation}
the field rotation relating the two bases also modifies the gauge charged-current couplings, generating an effective $W$ vertex involving the ALP:
\begin{equation}
	\mathcal{L}_{\rm gauge}\supset \frac{i a}{2 f_a} \frac{g}{\sqrt{2}} W_\alpha
	\bar \ell_i \left( C_{ij}^V - C_{ij}^A - C_{ij}^\nu \right)
	\gamma^\alpha P_L \nu_j\,. \label{eq:Wvertex}
\end{equation}
Unsuppressed by lepton masses, these interactions can induce sizeable effects when 
$C_{ij}^V - C_{ij}^A - C_{ij}^\nu \neq 0$. While we show that this is possible 
within an effective framework, in Appendix \ref{app:weak} we argue that it is difficult 
to deviate sizeably from the $SU(2)_L$-invariant limit in UV-complete theories without 
clashing with other experimental bounds. 

When the ALP is sufficiently light, the flavour off-diagonal couplings in Eq.~(\ref{eq:LagrangianCLepton}) can induce LFV decays of charged leptons with an ALP in the final state. In the limit $m_{\ell_i} \ll m_{\ell_j}$, the corresponding decay width is
\begin{equation}
		\Gamma( \ell_j\to \ell_i a)
		=
		\frac{m_j^3}{64\pi f_a^2}
		\left(|C^V_{ij}|^2+|C^A_{ij}|^2\right)
		\left(1-\frac{m_a^2}{m_{\ell_j}^2}\right)^2 \,,
\end{equation}
and the total decay width of a leptophilic ALP reads
\begin{equation}\label{eq:ALPdecwid}
		\Gamma_{a}(m_a)
		= \Gamma(a \to \gamma\gamma)
		+ \sum_{i,j} \Gamma(a \to \ell_i \ell_j)
		+ \sum_{i,j} \Gamma(a \to \nu_i \nu_j)\,,
\end{equation}
where the sum over the leptons runs over all kinematically accessible channels. The decay width into neutrinos is typically suppressed by the tiny neutrino masses and thus negligible while, even for a purely leptophilic ALP, the decay into two photons can still be generated via lepton loops. Following the notation of Ref.~\cite{Calibbi:2020jvd}, the partial decay width into a pair of charged leptons can be written as
\begin{equation}
	\Gamma(a \to \ell_i \ell_j)
	=
	\frac{m_a}{32\pi f_a^2}
	\left[
	(m_{\ell_i}-m_{ \ell_j})^2 |C_{ij}^{V}|^2 z_{+}
	+
	(m_{\ell_i}+m_{ \ell_j})^2 |C_{ij}^{A}|^2 z_{-}
	\right]
	\sqrt{z_{+}z_{-}} \, ,
\end{equation}
where $z_{\pm}\equiv 1-(m_{\ell_i}\pm m_{\ell_j})^2/m_a^2$. The two-photon decay width is instead given by
\begin{equation}
	\Gamma(a \to \gamma\gamma)
	=
	\frac{\alpha_{\rm em}^2\,E_{\rm eff}^2}{64\pi^3}\,
	\frac{m_a^3}{f_a^2}\,,
\end{equation}
where 
\begin{equation}
	E_{\rm eff} \equiv \sum_i C_{ii}^A \, B(\tau_i)\,,
	\qquad
	B(\tau) \equiv \tau \arctan^2\!\left(\frac{1}{\sqrt{\tau - 1}}\right) - 1 \,,
\end{equation}
and we have defined $\tau_i \equiv 4 m^2_{\ell_i}/m_a^2-i\epsilon$, with the sum running over the charged leptons. We ignore potential UV effects in the effective ALP photon coupling, e.g.,~originating from a $U(1)$ charge assignment with electromagnetic anomaly.

Light ALPs are typically long-lived and can therefore give rise to missing-energy signatures, which can be probed in a wide variety of experiments. In particular, for $m_a < m_\mu$, searches for exotic muon decays provide extremely sensitive probes of $e\mu$ ALP couplings, reaching very large values of $f_a$, for which all ALP couplings are suppressed making it long-lived, so that it escapes detection and is effectively invisible~\cite{GarciaiTormo:2011cyr, Uesaka:2020okd, Panci:2022wlc, Jho:2022snj, Xing:2022rob, Hill:2023dym, Knapen:2023zgi,  Bigaran:2025uzn, Calibbi:2020jvd}. For lower values of $f_a$, ALPs produced in muon decays can also be probed when their decay products are visible, such as an electron-positron or photon pair~\cite{Echenard:2014lma, Heeck:2017xmg, Hostert:2023gpk, Knapen:2023iwg, Fox:2024kda, Knapen:2024fvh}, and even via more exotic final states with high-multiplicity~\cite{Greljo:2025ljr}. 

The impressive sensitivity of these searches largely stems from the availability of extremely intense muon beams, which makes it possible to probe rare muon decays with very small branching ratios when backgrounds are under control. However, once the ALP mass exceeds the muon mass threshold, all bounds that rely on production in muon decays disappear, and significantly larger values of $C^{A,V}_{e\mu}/f_a$ are allowed.
For $m_a > m_\mu + m_e$, the decay channel $a \to e \mu$ opens up. In combination with the larger allowed couplings, this implies that the ALP typically decays promptly in this region of the parameter space, also due to the relatively unsuppressed coupling proportional to the muon mass. Consequently, different production mechanisms and search strategies must be considered to probe the ALP in this mass range.
As mentioned in the introduction, in this work we focus mostly on ALP couplings to first- and second-generation leptons, allowing for lepton flavour violation, and investigate the most promising production and detection mechanisms in this challenging region of parameter space.

Furthermore, we consider also some scenarios where the ALP has LFV
couplings to the $\tau$ lepton, providing an additional production channel via LFV 
$\tau$ decays. In the mass range where this is permitted, the decays $a\to \mu e,\, \mu\mu$ 
remain the dominant kinematically accessible visible channels. Consequently, the 
subsequent ALP decay can probe the underlying $e \mu$ couplings, allowing $\tau$ 
decays to provide indirect sensitivity to first and second generation ALP
interactions.

\section{ALP production in LFV decays}
\label{sec:production}

\begin{figure}[t]
\centering











\includegraphics[width=0.6\textwidth]{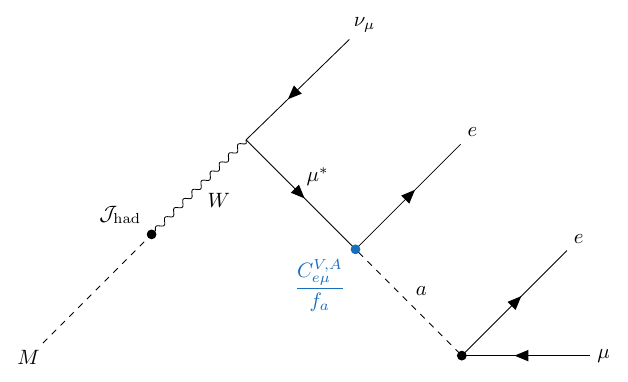}
\caption{\footnotesize Example of a meson decay mediated by LFV ALP emission. The internal muon is virtual, and its decay into an ALP is mediated by the $C_{e\mu}^{V,A}/f_a$ couplings. The ALP is instead produced on-shell and promptly decays into a dilepton final state.}\label{fig:mesondecayALP}
\end{figure}

In the presence of LFV couplings, ALPs can be produced in a variety of processes involving charged leptons. We focus here on the mechanisms most relevant for our analysis: decays of mesons, electroweak gauge bosons, and the $\tau$ lepton.

As already mentioned, we focus on ALPs with masses $m_a > m_\mu + m_e$, such that the decay $a \to e \mu$ is kinematically allowed and typically prompt. The relevant interactions arise from derivative ALP couplings to leptons $C^{V,A}_{ij}$, see Eq.~\eqref{eq:LagrangianCLepton}. In the following, we focus on couplings between the first two generations, i.e., $i,j=e,\mu$.

In weak decays the ALP can be emitted from the charged lepton {( or neutrino)} line, leading to three-body processes where the LFV structure is carried by the ALP vertex. These processes are particularly attractive experimentally because the subsequent decay $a \to e\mu$ produces distinctive multi-lepton final states with very small SM backgrounds. In particular, as we will discuss in detail in the following, one can always select the charge combination of the $e\mu$ pair such that the search is virtually background-free.

\subsection{Charged meson decays}

We first consider ALP production in charged pseudoscalar meson decays of the form
\begin{equation}
M^- \to e^-\, \bar{\nu}_\mu\, a \, ,
\end{equation}
where $M$ denotes a pseudoscalar meson such as $K$, $D$ or $B$. The ALP is emitted from the virtual charged lepton { or neutrino} line in the weak decay $M^- \to (\mu^-)^* \bar\nu_\mu$ { or $M^- \to (\bar{\nu}_e)^* e^-$}, followed by the LFV transition $(\mu^-)^* \to e^- a$ or { $(\bar{\nu}_e)^* \to \bar{\nu}_\mu a$}. The ALP will therefore decay back into SM particles according to its decay width given in Eq.~\eqref{eq:ALPdecwid}. For an example of a full decay chain with the ALP decaying into charged leptons, see \fref{fig:mesondecayALP}.

At energies well below the electroweak scale, the weak interaction is described by the effective Fermi operator
\begin{equation}
2\sqrt{2}\, G_F V_{ij}\,
(\bar u_i\gamma^\alpha P_L d_j)
(\bar\mu\gamma_\alpha P_L \nu_\mu)
+ {\rm h.c.}\,,
\end{equation}
where $V_{ij}$ is the CKM matrix element. The hadronic matrix element is parameterised by the meson decay constant,
\begin{equation}
\langle 0 | \bar u_i \gamma^\alpha\gamma_5 d_j | M(p)\rangle
= i f_M p^\alpha \,.
\end{equation}

Neglecting the electron mass and expanding to leading order in $m_\mu^2/m_M^2$, the decay width can be written as 
\begin{equation}
\Gamma(M^- \to e^- \bar\nu_\mu a)
\simeq
\frac{f_M^2 G_F^2 |V_{ij}|^2 m_M^5}
{3072\pi^3 f_a^2}
\left[
|C_L|^2\, g_0(x)
+(8|C_L|^2+{4{\rm Re} (C_L C^{\nu\star}_{e\mu}})+4|C_R|^2)
\left(\frac{m_\mu}{m_M}\right)^2
g_1(x)
\right]\,,
\label{eq:mesonrate}
\end{equation}
 where $x=m_a^2/m_M^2$ and we defined
\begin{equation}
C_L\equiv\frac{C^V_{e\mu}-C^A_{e\mu}-{ C^\nu_{e\mu}}}{2}\,,
\qquad
C_R \equiv \frac{C^V_{e\mu}+C^A_{e\mu}}{2}\,.
\end{equation}
{ The neutrino coupling $C^\nu_{e\mu}$ is defined in Eq.~(\ref{eq:ALPneutrinos}).}
The functions $g_0(x)$ and $g_1(x)$ arise from the three-body phase-space integration and read
\begin{align}
g_0(x) &= 1-8x+8x^3-x^4-12x^2\log x \,,\\
g_1(x) &= 1+9x-9x^2-x^3+6(x+x^2)\log x \,.
\end{align}

Wherever a comparison can be done, these expressions reproduce the results obtained in the lepton flavour-conserving case when the flavour indices are chosen accordingly, see Ref.~\cite{Altmannshofer:2022ckw}. In the $SU(2)_L$-invariant limit, where the LH charged-lepton coupling is aligned with the neutrino one, the contribution unsuppressed by lepton masses cancels and the decay becomes chirally suppressed ({note that, in the SU(2)-conserving case with exclusively left-handed ALPs, the rate likewise vanishes at order $m_\mu^2/m_M^2$}). The term proportional to $|C_L|^2$ in Eq.~\eqref{eq:mesonrate} therefore effectively parametrises the departure from this limit. This feature will play an important role in the phenomenological discussion below: in the exact $SU(2)_L$ limit the meson rates are strongly suppressed, while sizeable weak-breaking effects can substantially enhance the reach of both collider and fixed-target searches.

Among the meson channels, charged kaons and $D_s$ mesons are the most relevant for our analysis. At proton fixed-target experiments, the large kaon flux makes the decay $K\to e \nu_\mu a$ particularly well suited to probe promptly decaying ALPs through the subsequent LFV decay $a\to e\mu$. In contrast, at future high-luminosity $e^+e^-$ colliders running on the $Z$ pole, the very large number of produced heavy-flavour mesons implies that $D_s$ decays can become especially competitive. Although other charged mesons, such as $D$, $B$ or $B_c$, can also contribute in principle, their relevance is reduced either by CKM suppression or by the smaller number of mesons produced.

The corresponding experimental signature is highly distinctive. For prompt ALP decays, the process $M^- \to e^- \bar\nu_\mu a\, (\to e^- \mu^+)$ leads to a trilepton final state with same-charge electrons, missing energy and with a resonant $e\mu$ pair reconstructing the ALP mass. Since the charged leptons in the final state violate flavour conservation irrespective of the neutrino flavour assignment, there is no irreducible SM background. 

\subsection{Gauge boson decays}

ALPs can also be produced in decays of electroweak gauge bosons. The simplest example is
\begin{equation}
W^- \to e^-\, \bar{\nu}_\mu\, a \,,
\end{equation}
which proceeds through the same topology as the meson decay discussed above, with the ALP emitted from the virtual lepton line, and eventually decaying back into SM particles.

In the limit $m_W \gg m_a$, the decay width can be approximated as
\begin{equation}
\Gamma(W^- \to e^- \bar{\nu}_\mu a)
\simeq
\frac{g^2 m_W^3}{4096\pi^3 f_a^2}
\left[
|C_L|^2
+
\frac{4}{9}(2|C_L|^2+{{\rm Re} (C_L C^{\nu\star}_{e\mu}})+|C_R|^2)
\left(\frac{m_\mu}{m_W}\right)^2
g_2\!\left(\frac{m_a^2}{m_W^2}\right)
\right]\,,
\end{equation}
where
\begin{equation}
g_2(x)
=
-6(3x+1)\log x
-
(17-9x-9x^2+x^3) \,.
\end{equation}

\begin{figure}[t]
\centering











\includegraphics[width=0.49\textwidth]{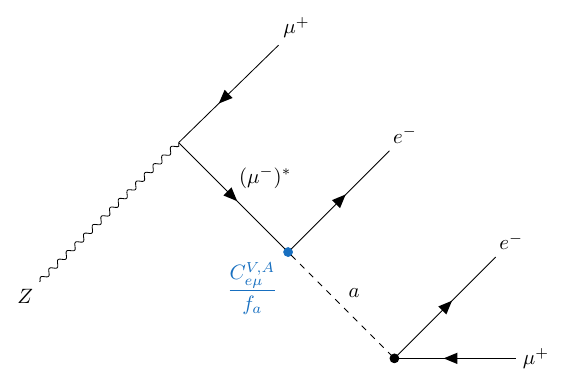}
\includegraphics[width=0.49\textwidth]{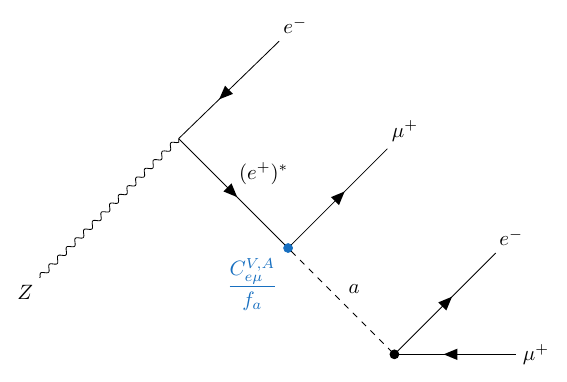}
\caption{\footnotesize Examples of Z boson decays mediated by LFV ALP emission. The internal charged lepton is virtual, and its decay into an ALP is mediated by the $C_{e\mu}^{V,A}/f_a$ couplings. The ALP is instead on-shell, promptly decaying into a dilepton final state.}\label{fig:bosondecayALP}
\end{figure}

A second important class of channels is provided by neutral-current decays such as
\begin{equation}
Z \to e^\pm \mu^\mp a \,,
\end{equation}
which proceed through analogous diagrams with the ALP radiated from a virtual charged lepton, before decaying again in SM particles. { The diagrams for these decays are} shown in \fref{fig:bosondecayALP}, yielding two same-sign electrons and two same-sign muons of opposite charge {(in the example $e^- e^- \mu^+ \mu^+$)}. 
Neglecting the electron mass and taking the leading term in the limit $m_\mu\ll M_Z$, the decay rate is\,\footnote{The couplings of the $Z$ boson to leptons are defined as
$-g/(2c_W) Z_\mu \left(g^V_\ell J^\mu_{\ell,V} + g^A_\ell J^\mu_{\ell,A}\right)$,
where $c_W$ denotes the cosine of the Weinberg angle, and $J^\mu_{\ell,V}$ and $J^\mu_{\ell,A}$ are the lepton vector and axial currents, respectively. In the analytical expression presented here, we neglect ALP emission from the vector current since the corresponding $Z$ coupling is accidentally suppressed, $g^{V}_\ell = -1/2 + 2s_W^2 \simeq 0.05$, compared with the axial coupling, $g^{A}_\ell = 1/2$. In the numerical results this contribution is included, together with the full dependence on $m_\mu$.}
\begin{equation}
    \Gamma(Z\to e^{\pm}\mu^\mp a)\simeq \frac{g^2 m_\mu^2 M_Z}{36864\pi^3 c_W^2 f_a^2}\left[\left|C_{e \mu}^{A}\right|^2+\left|C^{V}_{e \mu}\right|^2\right]\ g_3(x)\,,
    \label{eq:Zdecay}
\end{equation}
where $x=m^2_a/M_Z^2$ and 
\begin{align}
    g_3(x)= ~&
-x^3+27x^2-15x-11
-(9x^2+24x+3)\log x \nonumber\\
&-3x^2 \log^2 x
+6x^2 \log x\,\log(1+x)
+6x^2\left[
\operatorname{Li}_2\left(\frac{1}{1+x}\right)
-\operatorname{Li}_2\!\left(\frac{x}{1+x}\right)
\right],
\end{align}
which in the limit $m_a\ll M_Z$ can be approximated by 
\begin{equation}
    g_3(x\ll 1)\simeq -3\log x-11\,.
\end{equation}
Note that in the case of the $Z$ the rate remains chirally suppressed by the muon mass even in the presence of $SU(2)_L$-breaking couplings, because the effective $W$ vertex in Eq.~(\ref{eq:Wvertex}) does not contribute to this process.

For $W$ decays, the observable process is $W^- \to e^- \bar\nu_\mu a\,(\to e^- \mu^+)$, yielding two same-sign electrons, an oppositely charged muon, and missing energy. For $Z$ decays, instead, one obtains $Z\to e^\pm \mu^\mp a\,(\to e^\pm \mu^\mp)$, corresponding to four charged leptons with a resonant $e\mu$ pair. 
In both cases there is no irreducible SM background with the same flavour structure, so any residual contamination is expected to arise from rare processes combined with misidentification or accidental overlaps.
Future high-luminosity $e^+e^-$ colliders therefore provide the ideal environment to exploit these modes. At the $Z$ pole, FCC-ee and CEPC are expected to produce $\mathcal{O}(10^{12})$ $Z$ bosons, while dedicated runs at higher centre-of-mass energies can also yield very large samples of $W$ bosons. Even though the branching ratios for $W^- \to e^- \bar\nu_\mu a$ and $Z\to e^\pm\mu^\mp a$ are suppressed by $f_a^{-2}$, the enormous event statistics and the clean environment of a lepton collider make these channels sensitive probes of LFV ALP interactions. In particular, the $Z$ channel is especially attractive because it gives rise to a fully visible final state once the ALP decays promptly.

\subsection{Quarkonium decays}
\label{subsec:quarkonium}
Another interesting class of production channels is provided by heavy vector quarkonia decays,
\begin{equation}
	V \to e^\pm \mu^\mp a \,,
	\qquad
\text{with}
\qquad
	V = J/\psi,\ \Upsilon(nS)\,,
\end{equation}
where the ALP is emitted from a virtual charged lepton line. The corresponding topology is therefore analogous to that of $Z\to e^\pm\mu^\mp a$, with the neutral gauge boson replaced by a quarkonium resonance. These processes are dominated by  the annihilation of a neutral vector $c\bar c$ or $b\bar b$ bound state into a lepton pair via an off-shell photon, hence the non-perturbative QCD dynamics is encoded just by the vector quarkonium decay constant, defined as
\begin{equation}
	\langle 0 | \bar Q \gamma^\alpha Q | V(p,\varepsilon)\rangle
	=
	f_V m_V \varepsilon^\alpha \,,
\end{equation}
where $Q=c,b$ for charmonium and bottomonium, respectively, and $\varepsilon^\alpha$ denotes the polarisation vector of the quarkonium state. 
The decay $V\to e^\pm\mu^\mp a$ then proceeds through the effective coupling to an off-shell photon that decays into a lepton pair, followed by the LFV emission of the ALP from an off-shell muon. At the amplitude level, one may write schematically
\begin{equation}
	\mathcal{M}(V\to e^\pm \mu^\mp  a)
	\propto
	\frac{e^2 Q_Q f_V}{m_V}
	\varepsilon_\alpha
	\bar u_\mu \mathcal{T}^\alpha v_e\,,
\end{equation}
where $\mathcal{T}^\alpha$ contains the two diagrams in which the ALP is radiated from either lepton line ({electron or muon}), together with the corresponding LFV couplings $C^{V,A}_{e\mu}/f_a$. The resulting decay width is thus similar to that of $Z\to e \mu a$ in Eq.~(\ref{eq:Zdecay}). In the limit $m_\mu\ll m_V$, it reads 
\begin{equation}
    \Gamma(V\to e^{\pm}\mu^\mp a)\simeq \frac{\alpha^2 f_V^2 Q^2_Q m_\mu^2}{48\pi\, m_V f_a^2}\left[\left|C_{e \mu}^{A}\right|^2+\left|C^{V}_{e \mu}\right|^2\right]\ g_4(x)\,,
	\label{eq:quarkoniumrate}
\end{equation}
where {$x = m_a^2 / m_V^2$} and
\begin{align}
    g_4(x)=~& x^2 \log^2 x
+3x^2 \log x
-6x^2
-2x^2 \log x\,\log(1+x)\nonumber\\
&+2x^2\!\left[
\operatorname{Li}_2\!\left(\frac{x}{1+x}\right)
-\operatorname{Li}_2\!\left(\frac{1}{1+x}\right)
\right]
+8x+2x\log x-\log x-2\,,
\end{align}
and $\alpha=e^2/(4\pi)$. As in the case of the $Z$, the promising search strategy is to look for the striking signature with two same-sign electrons and two same-sign muons,
$V \to e^\pm \mu^\mp a\,(\to e^\pm \mu^\mp)$.
These essentially background-free searches are especially appealing in view of the large resonance samples expected at future lepton colliders. For charmonium, the proposed Super Tau Charm Factory (STCF)~\cite{Achasov:2023gey} is expected to collect an impressive $J/\psi$ data set, with annual yields of a few $10^{12}$ events and a total sample of $\mathcal{O}(10^{13})$ over the lifetime of the experiment. For bottomonium, the lower $\Upsilon(nS)$ resonances with $n\leq 3$ are also interesting, since they lie below the $\bar{B}B$ threshold and have sizeable leptonic branching fractions. Dedicated runs at the corresponding centre-of-mass energies at $B$ factories have already yielded samples of $\mathcal{O}(10^7-10^8)$ $\Upsilon(1S,2S,3S)$ \cite{Belle:2025pey, BaBar:2010vxb}, and runs at Belle~II could increase these datasets to $\mathcal{O}(10^{8}-10^{9})$ events \cite{Belle-II:2018jsg}. Nevertheless, as we will discuss in Section~\ref{sec:pheno}, the larger $J/\psi$ sample expected at STCF leads overall to a more promising projected sensitivity to the LFV ALP scenarios considered here.

\subsection{Tau decays}

\begin{figure}[t]
\centering

  






\includegraphics[width=0.5\textwidth]{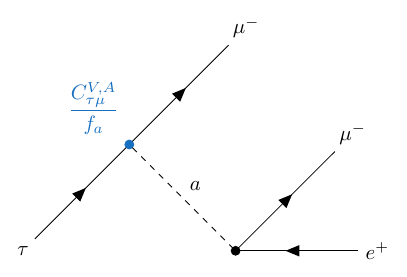}
\caption{\footnotesize Example of a $\tau$ decay mediated by LFV ALP emission. The ALP is on-shell, promptly decaying into a dilepton final state.}\label{fig:taudecayALP}
\end{figure}

Finally, ALPs can also be produced in LFV decays of the $\tau$ lepton,
\begin{equation}
\tau^- \to \ell^- a \,,
\qquad
\text{with}
\qquad
\ell=e,\mu\,,
\end{equation}
whose decay width takes the simple form
\begin{equation}
\Gamma(\tau \to \ell a)
=
\frac{m_\tau^3}{64\pi f_a^2}
\left(|C^V_{\ell\tau}|^2+|C^A_{\ell\tau}|^2\right)
\left(1-\frac{m_a^2}{m_\tau^2}\right)^2 \,.
\end{equation}

While these decays have mostly been studied as direct probes of LFV ALP couplings in the regime $m_a<m_\tau$, see Refs.~\cite{Calibbi:2020jvd,Cornella:2019uxs,Haghighat:2021djz,Bauer:2021mvw,Calibbi:2024rcm,Ema:2025bww}, in our setup they also provide an additional production mechanism for an  ALP that subsequently decays through the $e \mu$ couplings relevant to this work. This is particularly relevant in the mass window $m_\mu \lesssim m_a \lesssim m_\tau$, where the ALP can be produced on shell in the two-body decay of the $\tau$ and then decay visibly into two leptons, most notably through $a\to e\mu$ as shown in \fref{fig:taudecayALP} and, in the presence of flavour-diagonal muon couplings, $a\to\mu\mu$.

The resulting phenomenology depends on the ALP lifetime. If the ALP is long lived, the process can appear as $\tau\to \ell+\text{invisible}$, while for shorter lifetimes one obtains either prompt or displaced dilepton resonances in association with the primary charged lepton from the $\tau$ decay. In this way, $\tau$ decays provide a complementary handle on first- and second-generation ALP couplings even when production is driven by LFV interactions involving the third generation. High-statistics facilities such as Belle~II~\cite{Belle-II:2025urb} and future $Z$ factories~\cite{FCC:2025lpp, CEPCStudyGroup:2023quu} are therefore especially well suited to probe this class of scenarios.

\section{Phenomenological discussion}
\label{sec:pheno}

In this section, we bring together the production channels introduced in Section~\ref{sec:production} to determine the low-energy constraints on the ALP couplings to $e\mu$ and the projected sensitivities of future $e^+e^-$ colliders, Belle~II, STCF, and proton fixed-target or beam-dump experiments. Our analysis focuses on the region of parameter space with $m_a > m_\mu$.

\subsection{Low-energy constraints}\label{ssec:lowenergy}
In the absence of sizeable flavour-diagonal muon couplings, the dominant existing constraint on $e \mu$ ALP couplings arises from muonium--antimuonium oscillation\,---\,see Ref.~\cite{Chen:2026tdg} for a recent review\,---\,that probes a double insertion of the $e\mu$ coupling. For $m_a \gg m_\mu$, the conversion probability is given by \cite{Endo:2020mev, Bauer:2021mvw}
\begin{equation}
	\begin{aligned}
		P_{\text{Mu}\,\overline{\text{Mu}}} \;=\;
		\frac{m_{\mu}^4}{2\pi^2 a_0^6 \, \Gamma_{\mu}^2 m_a^4 f_a^4}
		\Bigg[
		&|c_{0,0}|^2 \left| (C^V_{e\mu})^2 - \left( 1 + \frac{1}{\sqrt{1+X^2}} \right) (C^A_{e\mu})^2 \right|^2
		\;+\; \\
		&|c_{1,0}|^2 \left| (C^V_{e\mu})^2 - \left( 1 - \frac{1}{\sqrt{1+X^2}} \right) (C^A_{e\mu})^2 \right|^2
		\Bigg]\, ,\label{eq:muonium}
	\end{aligned}
\end{equation}
where $\Gamma_\mu \simeq 3.0\times10^{-19}~\text{GeV}$ denotes the muon decay rate and $a_0 \simeq 2.69\times10^{5}~\text{GeV}^{-1}$ is the muonium Bohr radius. The parameter $X = 6.31\,(B/1~\text{T})$ is related to the external magnetic field $B$ employed in the experimental apparatus. The magnetic field also affects the probability of populating the initial hyperfine state labelled by the total angular momentum $(J,m_J)$, encoded in the coefficients $|c_{J,m_J}|^2$.
The most precise search for muonium--antimuonium oscillations to date was performed by the MACS experiment~\cite{Willmann:1998gd}, which operated with a magnetic field of $B = 0.1~\text{T}$. For this field strength, the relevant angular momentum factors entering Eq.~(\ref{eq:muonium}) are $|c_{0,0}|^2 = 0.32$ and $|c_{1,0}|^2 = 0.18$. MACS reported a $90\%$ confidence level upper bound on the oscillation probability of $P_{\text{Mu}\,\overline{\text{Mu}}} < 8.3\times10^{-11}$. The proposed MACE experiment, which plans to operate with the same magnetic field, aims to reach a sensitivity up to $P_{\text{Mu}\,\overline{\text{Mu}}} < 7\times10^{-14}$ \cite{Bai:2024skk,Chen:2026tdg}, three orders of magnitude better than the current one.

The situation changes qualitatively when the flavour-conserving coupling $C^A_{\mu\mu}$ is also present. In that case, muon-ALP loop diagrams involving both $C_{e\mu}^{V,A}$ and $C_{\mu\mu}^A$ generate the radiative decay $\mu\to e\gamma$, which can provide stringent bounds on the parameter space. Following Ref.~\cite{Cornella:2019uxs}, the rate can be written as

\begin{equation}
	\Gamma(\mu\to e\gamma) = \frac{\alpha\, m_\mu^3}{2 (64\pi^2) ^2 f_a^4} 
	\,\left|C_{\mu\mu}^A\right|^2\left[
	\left|C_{e\mu}^A\right|^2  + \left|C_{e\mu}^V\right|^2
	\right]\,h^2(m_a^2 / m_\mu^2)\,,\label{eq:mutoegamma}
\end{equation}
where
\begin{align}
	h(x) = \frac{(x - 3)\,x^2 \log x}{x - 1} - 2x + 1 - 
	2 \sqrt{x - 4}\, x^{3/2} \log\!\left( \frac{\sqrt{x - 4} + \sqrt{x}}{2} \right)\,.
\end{align}
Stringent limits on the effective couplings contributing to $\mu \to e \gamma$ can also be derived from future searches for $\mu \to e$ conversion in nuclei and $\mu \to eee$. Virtual ALP exchange can mediate $\mu\to eee$ already at tree level when $C^A_{ee}\neq 0$, but this contribution is suppressed by the electron mass, that is, by a factor $(m_e/m_\mu)^2$ relative to the above one. If $\mu\to eee$ is instead dominated by an off-shell photon $\mu\to e\gamma^*$ that decays into an electron-positron pair, the ratio between ${\rm BR}(\mu\to e\gamma)$ and ${\rm BR}(\mu\to eee)$ is fixed~\cite{Kuno:1999jp}:
\begin{equation}
	\frac{{\rm BR}(\mu\to eee)}{{\rm BR}(\mu\to e\gamma)}
	\simeq \frac{\alpha}{3\pi}\left[\log\!\left(\frac{m_\mu^2}{m_e^2}\right)-\frac{11}{4}\right]
	\simeq 6.1\times 10^{-3}\,.
\end{equation}
The projected Mu3e sensitivity, ${\rm BR}(\mu\to eee)\lesssim 10^{-16}$ \cite{Mu3e:2020gyw}, then would translate into the indirect bound
${\rm BR}(\mu\to e\gamma)\lesssim 1.6\times 10^{-14}$,
which is stronger than the expected final sensitivity of MEG~II, ${\rm BR}(\mu \to e \gamma) < 6 \times 10^{-14}$~\cite{Moritsu:2022lem}.  Likewise, when $\mu \to e$ conversion in nuclei is dominated by the dipole transition, the conversion rate (CR) is correlated to ${\rm BR}(\mu \to e \gamma)$. For an aluminium target, one finds ${\rm BR}(\mu \to e \gamma)/{\rm CR}(\mu\,{\rm Al} \to e\,{\rm Al}) \simeq 389$~\cite{Kuno:1999jp,Kitano:2002mt}. Hence, the expected future limits of the upcoming Mu2e and COMET experiments, \mbox{${\rm CR}(\mu\,{\rm Al} \to e\,{\rm Al}) \lesssim 6\times 10^{-7}$}~\cite{Kuno:2013mha,Mu2e:2014fns}, will translate into the indirect bound
$ {\rm BR}(\mu \to e \gamma) \lesssim 2.3 \times 10^{-14}$, a sensitivity that is comparable to (but slightly weaker than) that expected from $\mu\to eee$.

\subsection{Future $e^+ e^-$ colliders}\label{subsec:colliders}

Future $e^+e^-$ colliders such as FCC-ee~\cite{FCC:2025lpp} and CEPC~\cite{Ai:2025cpj} provide ideal environments for the search strategy proposed here. This is due both to the extremely large expected samples of electroweak gauge bosons, tau leptons, and heavy-flavour mesons, and to the clean experimental environment of a lepton collider, which is particularly well suited for reconstructing final states containing neutrinos.

When operating at the $Z$ pole, future lepton colliders are expected to produce $4\text{--}5 \times 10^{12}$ $Z$ bosons \cite{Ai:2024nmn, Blondel:2021ema}. $Z$ decays at such ``Tera-$Z$ factories'' are therefore expected to yield samples of order $10^{11}$ $\tau$-lepton pairs and $D_s^{\pm}$ mesons~\cite{Ai:2024nmn}. A comparable number of $D$ and $B^{\pm}$ mesons is also expected. However, as shown in Eq.~\eqref{eq:mesonrate}, their decay rates into ALPs are suppressed by  $V^2_{cd}$ and $V^2_{ub}$, respectively, making these channels less competitive than searches involving $D_s$ decays. The decay $B_c^{\pm} \to e^\pm \nu_\mu a$ is instead only mildly suppressed by $V_{cb}$, but the expected number of $B_c^{\pm}$ mesons produced at Tera-$Z$ factories is roughly three orders of magnitude smaller than that of $D_s^{\pm}$ \cite{Ai:2024nmn}. It is worth mentioning that, during the planned operations at the $WW$ threshold and in the $\sqrt{s}=240\,\mathrm{GeV}$ Higgs factory stage, future lepton colliders are also expected to produce $\O{(10^{9})}$ $W$ bosons.

For charged mesons and $W$ bosons, the processes that can be investigated at these facilities are 
\begin{equation}
	\Gamma(M^-, W^- \to e^- e^- \mu^+ \bar{\nu}_\mu)
	= \Gamma(M^-, W^- \to e^- \bar{\nu}_\mu a) \times {\rm BR}(a \to e^- \mu^+)\, ,
    \label{eq:M/W}
\end{equation}
where the ALP is emitted from the charged lepton line and subsequently decays promptly into an $e\mu$ pair with an electron of the same charge as the decaying meson. Similarly, in the case of $Z$ decays, the relevant process is
\begin{equation}
	\Gamma(Z \to e^{\pm} e^{\pm} \mu^{\mp} \mu^{\mp})
	= \Gamma(Z \to e^{\pm} \mu^{\mp} a) \times {\rm BR}(a \to e^{\pm} \mu^{\mp})\, .
\end{equation}
In both cases, the overall lepton flavour violation in the process corresponds to $|\Delta L_e| = |\Delta L_\mu| = 2$, which can only be mimicked within the SM through a combination of rare processes\,---\,necessarily involving a larger number of neutrinos, hence more missing energy\,---\,and/or particle misidentifications. Notice that, although the neutrino flavour in Eq.~\eqref{eq:M/W} can not be observed, the final state containing three charged leptons violates lepton-flavour conservation for any possible neutrino flavour assignment and therefore cannot arise from SM processes. Moreover, the final states feature a well-defined invariant mass corresponding to the decaying particle, which provides an additional handle to suppress potential accidental backgrounds.

For example, processes such as 
$e^+ e^- \to e^- e^- e^+ e^+$,
$e^+ e^- \to \mu^- \mu^- \mu^+ \mu^+$, or
$e^+ e^- \to e^- \mu^- e^+ \mu^+$
can mimic the signal $Z \to e^{\pm} e^{\pm} \mu^{\mp} \mu^{\mp}$ only if two $e\leftrightarrow \mu$ misidentifications occur. Using \texttt{MadGraph5}~\cite{Maltoni:2002qb, Alwall:2011uj}, we find cross sections at the $Z$ pole of order
$\sigma_{4\ell}\simeq {\rm few}\times 10^{-3}\,\mathrm{pb}$.
For the projected integrated luminosity of Tera-$Z$ factories,
$\mathcal{L}\simeq100~\mathrm{ab}^{-1}$~\cite{CEPCStudyGroup:2023quu, FCC:2025uan},
this corresponds to
$N_{4\ell}\simeq {\rm few}\times 10^5$
events. Therefore, an $e\leftrightarrow \mu$ misidentification rate below $0.1\%$, which appears realistic~\cite{CEPCStudyGroup:2018ghi, FCC:2025lpp}, would already suppress this background to fewer than one expected event. Another possible background to the same search arises from
$e^+ e^- \to \tau^+ \tau^- \tau^+ \tau^-$,
which has a cross section comparable to that of the four-electron/muon processes. This channel can fake the signal if all taus decay leptonically, and if the charged leptons are produced close to the kinematic endpoints of their energy distributions. However, this background is suppressed by ${\rm BR}(\tau \to \ell \nu \nu)^4$, as well as by the requirement that the neutrinos carry only negligible missing energy/momentum within the detector resolution. We therefore expect it to be entirely irrelevant. For reference, a similar potential background to $Z\to e^\mp \mu^\pm$ searches arising from the much larger sample of $\approx10^{11}$ $Z\to \tau^+ \tau^-$ events, is estimated to be negligible~\cite{Dam:2018rfz}. {Regarding the meson decays considered here, potential backgrounds could arise from rare four-body semileptonic decays involving neutrinos, such as $D_s\to K^+ K^- e \nu$, combined with at least two simultaneous hadron-to-lepton misidentifications. We expect this background to be negligible as well, although a more precise estimate would require specifying the details of the experimental search strategy and event selection.}

For the above reasons, we assume the backgrounds to be negligible and estimate the sensitivity of future experiments based on their projected event statistics. The expected number of signal events is therefore
\begin{align}
	N^{M,W}_{\rm sig}&=\epsilon_\text{eff} \,N_{M,W} \times {\rm BR}(M^-,\ W^-\to e^- \bar{\nu}_\mu a)\times {\rm BR}(a\to e^- \mu^+)\nonumber \\ 
	N^Z_{\rm sig}&=\epsilon_\text{eff} \,N_Z \times {\rm BR}(Z\to e^\mp \mu^\pm a)\times {\rm BR}(a\to e^\mp \mu^\pm)\,, \label{eq:neventsZMW}
\end{align} 
where $\epsilon_\text{eff}$ accounts for detector efficiencies and geometrical acceptances. To estimate it, we implement the SM and ALP interactions in \texttt{MadGraph5} \cite{Maltoni:2002qb, Alwall:2011uj} and generate events for $e^+ e^- \to Z \to e^{\pm} e^{\pm} \mu^{\mp} \mu^{\mp}$, while we simulate a CEPC-like detector response using a dedicated \texttt{Delphes} card \cite{deFavereau:2013fsa}. We perform the simulation for $m_a=0.12,\, 0.5,\, 1,\, 5,\, 10,\, 50\,\mathrm{GeV}$, fixing $f_a=1\,\mathrm{TeV}$ and $C^A_{e\mu}=1$, with all other ALP couplings set to zero. We find that, in all cases, at least $\approx 50\%$ of the events are correctly reconstructed. We therefore adopt a conservative overall efficiency $\epsilon_\text{eff} \approx 0.5$ and assume a comparable performance for processes involving charged mesons and $W$ bosons. The projected $90\%$ CL limit on the ALP couplings from a given process is then obtained by requiring the expected number of signal events to satisfy $N_{\rm sig}=2.3$.
Searches in quarkonium decays can target final states analogous to those considered for the $Z$ boson, namely
$V \to e^\pm \mu^\mp (a \to e^\pm \mu^\mp)$, with an expected number of signal events given by
\begin{equation}
N_{\rm sig}^V=
\epsilon_{\rm eff}\, N_V \times {\rm BR}(V \to e^\mp \mu^\pm a)\times {\rm BR}(a \to e^\mp \mu^\pm)\, .
\end{equation}
where $V=J/\psi, \Upsilon(nS)$ and $N_V$ is the size of the quarkonium sample.
For fixed $C^{A,V}_{e\mu}/f_a$ and $m_a$, the branching ratio of $\Upsilon(nS)\to e\mu a$ for the $n<3$ $\Upsilon(nS)$ resonances is roughly one order of magnitude smaller than that of $J/\psi \to e\mu a$, owing to the larger quarkonium mass and the smaller electric charge of the $b$ quark (see Eq.~(\ref{eq:quarkoniumrate})). In addition, as we discuss in Section \ref{subsec:quarkonium} the number of $\Upsilon(nS)$ resonances expected to be collected at Belle~II is about $4$–$5$ orders of magnitude smaller than the $\approx10^{13}$ $J/\psi$ sample anticipated at STCF. We therefore restrict our analysis to $J/\psi$ decays, which provide a more promising probe.

\begin{figure}[t]
	\centering
	\includegraphics[width=0.7\textwidth]{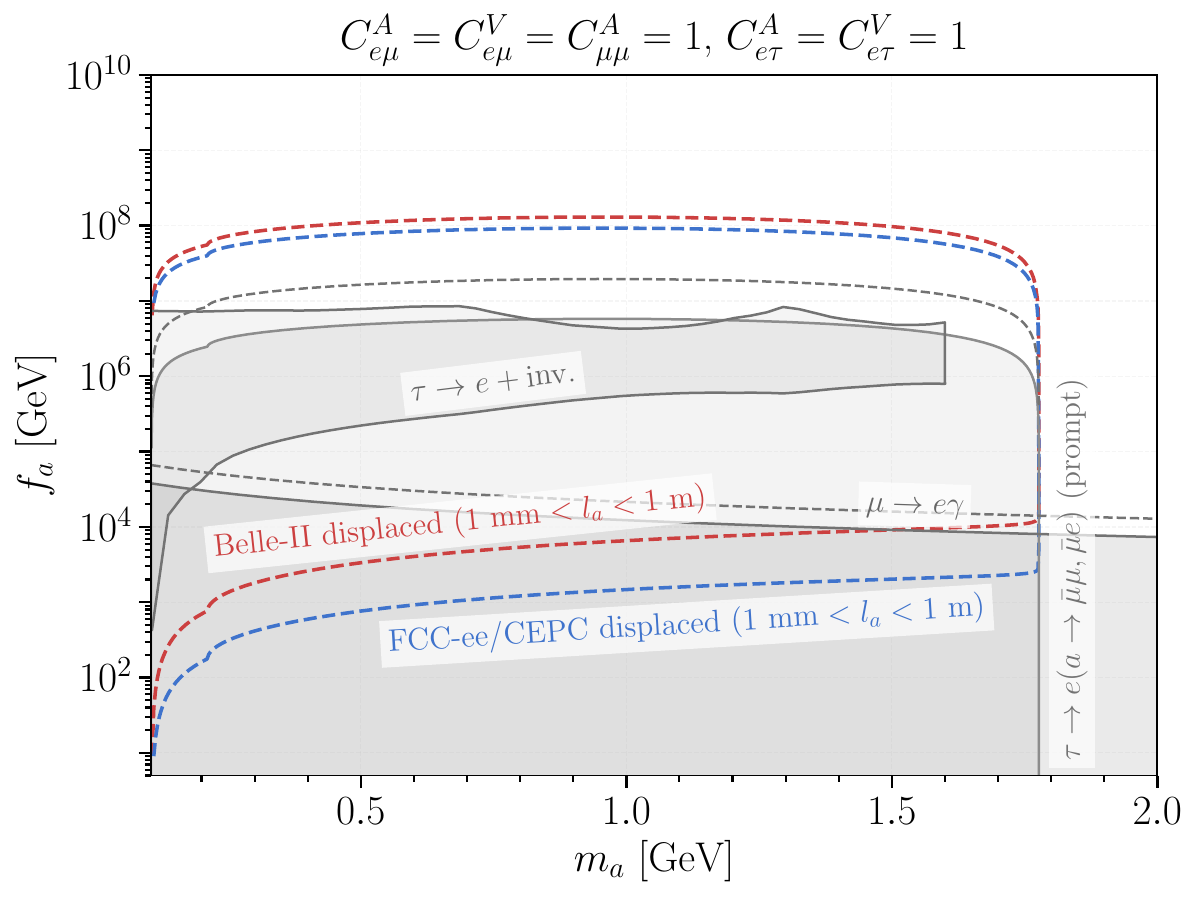}
	\caption{\footnotesize Projected sensitivities of Belle~II and the proposed Tera-$Z$ factories CEPC/FCC-ee to displaced-vertex signatures arising from ALP decays produced by $\tau \to e a$, assuming
		$|C_{\tau e}^{V,A}| = |C_{e\mu}^{V,A}| = |C_{\mu\mu}^{A}|$.
		The shaded grey region is excluded by current constraints. For prompt ALP decays, the relevant bounds are ${\rm BR}(\tau \to e \bar{\mu} \mu)<2.4\times 10^{-8}$ and ${\rm BR}(\tau \to e \bar{\mu} e)<1.5\times 10^{-8}$ \cite{Hayasaka:2010np, Belle-II:2025urb}. For ALPs decaying outside the detector, the constraint ${\rm BR}(\tau \to e a) \lesssim \text{few}\times10^{-4}$ \cite{Belle-II:2022heu, Belle:2025bpu} applies; the precise limit depends mildly on the ALP mass and extends up to $m_a=1.6$ GeV. The bound ${\rm BR}(\mu\to e \gamma)\lesssim 1.5\times 10^{-13}$ \cite{MEGII:2025gzr} constrains $f_a$ when both the $\mu\mu$ and $e\mu$ couplings are present as described by Eq.~(\ref{eq:mutoegamma}). The grey dashed lines correspond to the future projected sensitivities: ${\rm BR}(\tau \to e \bar{\mu} \ell)\lesssim 10^{-10}$ \cite{Banerjee:2022xuw} at Belle~II and the indirect limit  ${\rm BR}(\mu\to e \gamma)\lesssim 1.6\times 10^{-14}$ from the $\mu\to eee$ experiment Mu3e (see Section~\ref{ssec:lowenergy}). The regions enclosed by the red (Belle~II) and blue (CEPC/FCC-ee) dashed contours are accessible via the displaced-vertex searches for an ALP with a decay length $1~\mathrm{mm} < l_a < 1~\mathrm{m}$ in the lab frame. }
	\label{fig:displaced}
\end{figure}

A complementary production mechanism arises if the ALP also couples to the $\tau$ lepton. In this case the ALP can be produced in the two-body decay $\tau \to \ell a$, with $\ell=e,\mu$, which is induced by non-zero $C^{V,A}_{\ell\tau}$ in Eq.~(\ref{eq:LagrangianCLepton}).
Such a scenario can be probed both at Tera-$Z$ factories, where large samples of $\tau$ leptons are produced in $Z$ decays, and at flavour factories such as Belle~II. In the mass range of interest, $m_\mu \lesssim m_a \lesssim m_\tau$, the ALP subsequently decays predominantly via
\begin{equation}
	\Gamma_a \simeq 
	\Gamma(a \to e^\pm \mu^\mp) + \Gamma(a \to \mu^+\mu^-)\,,
\end{equation}
in the presence of non-vanishing $e\mu$ and $\mu\mu$ couplings. As a result, even when the ALP is produced in $\tau$ decays, its subsequent decay can still probe  $e\mu $ interactions. The resulting phenomenology depends on the ALP decay length $l_a = \beta_a \gamma_a\, \Gamma_a^{-1}$, where $\beta_a$ and $\gamma_a$ are the ALP velocity and boost in the lab frame. If the ALP is sufficiently long-lived to escape the detector, the observable signature reduces to $\tau \to \ell + \text{inv.}$, for which Belle and Belle~II currently set limits ${\rm BR}(\tau \to \ell a) \lesssim \text{few}\,\times10^{-4}$~\cite{Belle-II:2022heu, Belle:2025bpu}, depending on the value of $m_a$. In this regime, no direct information on the $e \mu$ couplings can be extracted.
If instead the ALP decays promptly, the signature becomes $\tau \to \ell a\,(\to \mu\mu)$ or $\tau \to \ell a \,(\to e \mu)$, producing a resonant dilepton pair. For realistic values of $f_a\gtrsim 1$ TeV,  the intrinsic ALP width is much smaller than the experimental mass resolution to be measured, so the individual couplings cannot be disentangled, although their ratio can in principle be probed through the relative rates of the two channels. On the other hand, when the ALP decays with a measurable displacement, displaced-vertex signatures arise. In this regime, the decay length directly probes the ALP lifetime and therefore the $e\mu$ and $\mu\mu$ couplings that control it. Dedicated searches for displaced dilepton vertices in $\tau$ decays at Belle~II or future $e^+e^-$ colliders could therefore explore an additional region of parameter space in the window $m_\mu \lesssim m_a \lesssim m_\tau$. 

In order to appreciate the sensitivity that can be achieved by exploiting displaced ALPs produced in $\tau$ decay, we assume that the ALP couples to $\tau e$, $e \mu$, and $\mu\mu$ (vector and axial) currents with comparable strength,
\begin{equation}
	|C_{\tau e}^{V,A}| \;=\; |C_{e\mu}^{V,A}| \;=\; |C_{\mu\mu}^{A}| \, ,
\end{equation}
so that the phenomenology is fully determined by the ALP mass $m_a$ and the decay constant $f_a$.  
In this scenario, the production rate is fixed by the two-body decay $\tau \to e a$, while the ALP decay length is controlled by its decays into $\mu\mu$ and $e\mu$, with the same partial widths.
We define the experimentally accessible region by requiring that the ALP decays within $1~\mathrm{mm} < l_a < 1~\mathrm{m}$ (i.e.,~inside the inner detector) so that the expected number of signal events is
\begin{equation}
	N_{\rm sig}
	=
	N_\tau \times {\rm BR}(\tau \to e a)
	\times	\left[\exp\!\left(\frac{-1~\mathrm{mm}}{l_a}\right)
	- \exp\!\left(-\frac{1~\mathrm{m}}{l_a}\right)\right] \,,
\end{equation}
where $N_\tau$ is the total number of $\tau$ leptons.\footnote{We take as the expected number of produced $\tau$ leptons $N_\tau = 4.5\times 10^{10}$ and $2\times 10^{11}$ for the full integrated luminosities of Belle~II and the Tera-$Z$ runs of CEPC/FCC-ee, respectively.} 
Assuming negligible background, we define the projected sensitivity by requiring $N_{\rm sig} \leq 2.3$,
which selects the region in the $(m_a, f_a)$ plane shown in \fref{fig:displaced}.
Current bounds for the LFV decays $\tau \to e \bar{\mu} \mu$ and $\tau \to e \bar{\mu} e$~\cite{Hayasaka:2010np, Belle-II:2025urb} constrain the complementary region of parameter space in which the ALP decays promptly, corresponding to decay lengths $l_a \lesssim 1~\mathrm{mm}$, while searches for $\tau \to \ell + \text{inv.}$ are sensitive to ALPs that escape the detector $l_a\gtrsim 1\,\mathrm{m}$.
As illustrated by \fref{fig:displaced}, displaced-vertex searches can extend beyond the reach of existing constraints. Although Belle~II is expected to collect roughly $\times4.5$ fewer $\tau$ pairs than future $e^+e^-$ colliders, the upper reach of the displaced-vertex sensitivity is slightly better at Belle~II. This is due to the difference in boosts between the two setups being compensated by the difference in sample size. In other words, the larger boost of Tera-$Z$ factories reduces the probability that the ALP decays inside the detector. 
Indeed, in the region of large $f_a$ (long-lived ALPs) the probability of decaying within $l_{\rm max}=1 $ m is $P_\text{decay}\simeq l_{\rm max}/l_a$, and the ratio between the expected number of events is hence $(N^{\rm TeraZ}_\tau/N^{\rm Belle~II}_\tau)\times (l^{\rm Belle~II}_a/l^{\rm TeraZ}_a)\simeq 0.54$, where we have used that $l^{\rm Belle~II}_a/l^{\rm TeraZ}_a=\gamma^{\rm Belle~II}_a/\gamma^{\rm TeraZ}_a$, with $\gamma^{\rm Belle~II}_a=3.1$ and $\gamma^{\rm TeraZ}_a=25.7$ being the average boosts in the two setups. 

\subsection{Proton fixed-target and beam-dump experiments}\label{subsec:beamdump}
Fixed-target and beam-dump experiments based on high-intensity proton beams provide an interesting opportunity to probe ALPs produced in meson decays, thanks to the very large yields of light and heavy mesons generated in target collisions. These facilities are therefore well suited to explore rare production modes such as those induced by the LFV couplings considered here.

The NA62 experiment at CERN is currently the leading facility for studies of rare kaon decays \cite{NA62:2017rwk}. In the present context, it provides the most relevant setting to search for the process $K \to e \nu_\mu a \,(a \to e \mu)$, which gives rise to the same characteristic LFV signature discussed in Section~\ref{subsec:colliders} when the ALP decay is prompt. As in the collider case discussed above, the SM background is expected to be negligible, up to accidental combinations and particle misidentification. The corresponding sensitivity can therefore be estimated from existing searches for rare LFV kaon decays, i.e., ${\rm BR}(K^- \to \bar{\nu}_\mu e^-  e^- \mu^+)\lesssim 10^{-11}$ \cite{Bician:2022lsm}.

Proton beam-dump experiments are instead sensitive to ALPs that are sufficiently long-lived to traverse a thick target and decay inside a downstream fiducial volume. In this case, ALPs can be produced in the collisions of the proton beam with the target, for instance through meson decays, and subsequently be observed via their visible decay products. Beam-dump facilities therefore probe a region of parameter space complementary to prompt-decay searches, extending the sensitivity to smaller couplings and longer lifetimes.

The production of ALPs in proton beam dumps has already been considered in the literature in scenarios with non-zero couplings to the tau lepton \cite{Ema:2025bww}, where the dominant source of ALPs are two-body $\tau$ decays, with the $\tau$ leptons themselves produced mainly in $D_s$ decays generated in the primary proton-target collisions. Here we instead focus on the case in which the ALP is produced directly in the three-body decay $M^- \to e^- \bar\nu_\mu a$  of Eq.~(\ref{eq:mesonrate}) induced by the $e \mu$ couplings. The parent meson $M$ can be either a $D_s$ or a $K$. We do not consider pions because the phase space available for $m_a>m_\mu$ is small. We simulate the production spectrum of the parent mesons in the target and sample the ALP kinematical distribution arising from the meson decay. The ALP must traverse the distance $L$ to the detector face and subsequently decay within the fiducial volume. The probability of this occurring is
\begin{equation}
	P_{a}= \exp \left( -\frac{\Gamma_a L}{\gamma_a \beta_a} \right) \left[ 1 - \exp \left( -\frac{\Gamma_a \lambda}{\gamma_a \beta_a} \right) \right]\,,
\end{equation}
where $\lambda$ is the path length inside the fiducial volume, $\Gamma_a$ is the ALP decay width, and $\gamma_a=E_a/m_a$ is the boost factor. Both $L$ and $\lambda$ also depend on the direction of the ALP velocity, which can be parametrized by a polar angle $\theta_a$ and an azimuthal angle $\phi_a$, identifying $z$ as the direction of the primary proton beam.  The total number of expected signal events is therefore given by:
\begin{equation}
	N_{\rm sig}= N_{M} \int \frac{1}{\Gamma_M} \frac{d\Gamma_{M \to e \nu_\mu a}}{dE_a d \theta_a d \phi_a} P_{a} \, dE_a d\theta_a d\phi_a\,,\label{eq:signalSHIP}
\end{equation}
where $N_M$ is the number of mesons produced and $\Gamma_M$ their total decay rate. We evaluate the integral  numerically by averaging the decay probability over the Monte Carlo samples. We give the details of the simulation we perform to obtain the ALP kinematical distribution in Appendix~\ref{app:beamdump}. Because the three-body production rate of ALPs is suppressed (especially in the $SU(2)_L$-conserving limit, see Eq.~(\ref{eq:mesonrate})) and because the ALP typically decays over relatively short length scales once the $e\mu$ and $\mu\mu$ channels are kinematically open, the proposed Search for Hidden Particles (SHiP) experiment at CERN appears to be the most promising to probe this class of scenarios using beam dumps~\cite{Albanese:2948477}. SHiP is designed to use the 400 GeV SPS proton beam and is expected to accumulate up to $6\times10^{20}$ protons on target. The fiducial volume begins about $34$ m downstream of the target and is $\approx 50~\mathrm{m}$ long with the shape of a pyramidal frustum, while the upstream and downstream transverse dimensions are approximately $1.0\times2.7~\mathrm{m}^2$ and $4\times6~\mathrm{m}^2$, respectively. Other proton beam-dump experiments typically have longer baselines, and the corresponding event yields we estimate are not competitive with the projected sensitivity of SHiP. We therefore focus on the latter.

Compared with the $D_s$ case, estimating the kaon contribution in a beam-dump setup such as SHiP is more involved. The reason is that $D_s$ mesons are very short-lived and decay promptly after production, whereas charged kaons can undergo multiple interactions in the target before decaying. As a result, only a subset of the produced kaons is relevant for the signal. In particular, the useful kaons are those produced in the target that decay in flight, before being stopped or absorbed, and that retain a sufficiently large forward momentum to point toward the downstream fiducial volume. Kaons that decay after coming to rest are not included, since their decay products are emitted isotropically and thus have a small probability of entering the SHiP decay volume. A similar analysis has been carried out for heavy neutral leptons produced in kaon decays at SHiP in Ref.~\cite{Gorbunov:2020rjx}.

\subsection{Combined results}

We consider two benchmark scenarios, distinguished by the relative size of the flavour-violating $e\mu$ coupling and the flavour-diagonal coupling to muons.

\begin{itemize}
	\item \textbf{Scenario A:}
	\begin{equation}
		C^{A}_{\mu\mu}\ll C^{V,A}_{e\mu}\,.
	\end{equation}
	This situation can arise accidentally in UV-complete models, or as a consequence of flavour symmetries. 
    An example is provided by the residual  $\mathbb Z_4$  flavour symmetry in the charged lepton sector considered in Ref.~\cite{Calibbi:2025fzi}. For instance, assigning flavour charge 0 (modulo 4) to electrons, and charge 2 to muons, $\tau$, and {the scalar field that $a$ belongs to}, only allows non-vanishing $C^{V,A}_{e\mu}$ and $C^{V,A}_{e\tau}$ and, in particular, forbids all flavour-conserving ALP interactions.\footnote{We are grateful to A.~Greljo for pointing out this possibility.} %
    Obviously, a different $\tau$ charge assignment would also forbid the LFV ALP-$\tau$ interactions. 
    ALPs within the context of $\mathbb Z_4$ and $\mathbb Z_8$ flavour symmetries have been considered in Ref.~\cite{Greljo:2024evt}.
	\item \textbf{Scenario B:}
	\begin{equation}
		C^{A}_{\mu\mu}\simeq C^{V,A}_{e\mu}\,.
	\end{equation}
	This scenario is realised within a broad class of models in which flavour-conserving and flavour-violating ALP couplings arise from flavour non-universal charge assignments of the underlying $U(1)$ symmetry and the rotations that diagonalise the lepton mass matrix in the $e\mu$ sector are $\O{(1)}$, see, e.g.,~Refs.~\cite{Calibbi:2020jvd,Panci:2022wlc}.
\end{itemize}

In both scenarios, allowing for a non-negligible coupling to electrons $C^A_{ee}$ would not qualitatively modify the discussion, because its effects are suppressed by the small electron mass.\footnote{In the presence of $SU(2)_L$-breaking effects, the unsuppressed $W$ vertex involving electrons does not contribute to the LFV processes discussed in this work, and could instead be probed with the strategies proposed in Ref.~\cite{Altmannshofer:2022ckw}. In our setup, the only observable that could be significantly affected by such effects is the ALP production rate at SHiP via $D_s$ and $K$ decays, which could be enhanced. Therefore, in the presence of non-negligible $ee$ couplings and weak-breaking, the limits derived here from LFV production should be regarded as conservative.}
Within each benchmark, we further distinguish between predominantly LH and right-handed (RH) ALP couplings. In the LH case, because we do not include the corresponding couplings to neutrinos, the charged-lepton interactions are effectively misaligned with the neutrino sector. As mentioned in Section~\ref{sec:production}, this implies a sensitivity to $SU(2)_L$-breaking effects that can induce the unsuppressed weak vertex of Eq.~(\ref{eq:Wvertex}), with important consequences for meson and $W$-boson decays. We discuss the challenges in accommodating large weak violation in a UV complete model in Appendix \ref{app:weak}. By contrast, in the RH case no analogous effect arises, and the resulting rates remain chirally suppressed.

\begin{figure}[!htp!]
	\centering
	\includegraphics[width=0.64\textwidth, trim=0 0 0.5cm 0.5cm, clip]{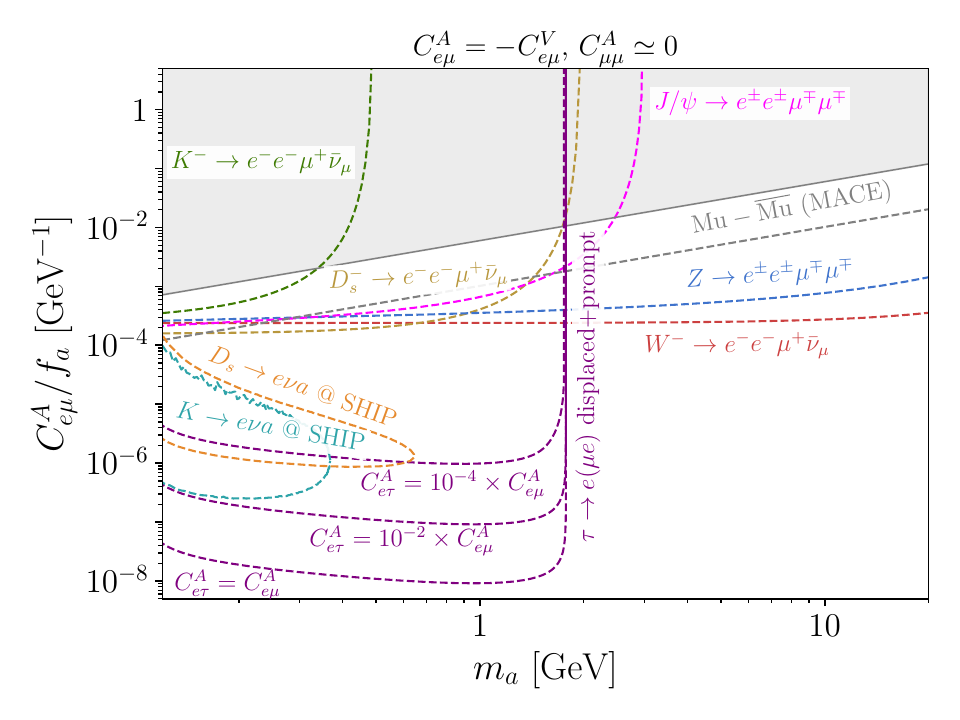}
	\includegraphics[width=0.64\textwidth, trim=0 0 0.5cm 0.5cm, clip]{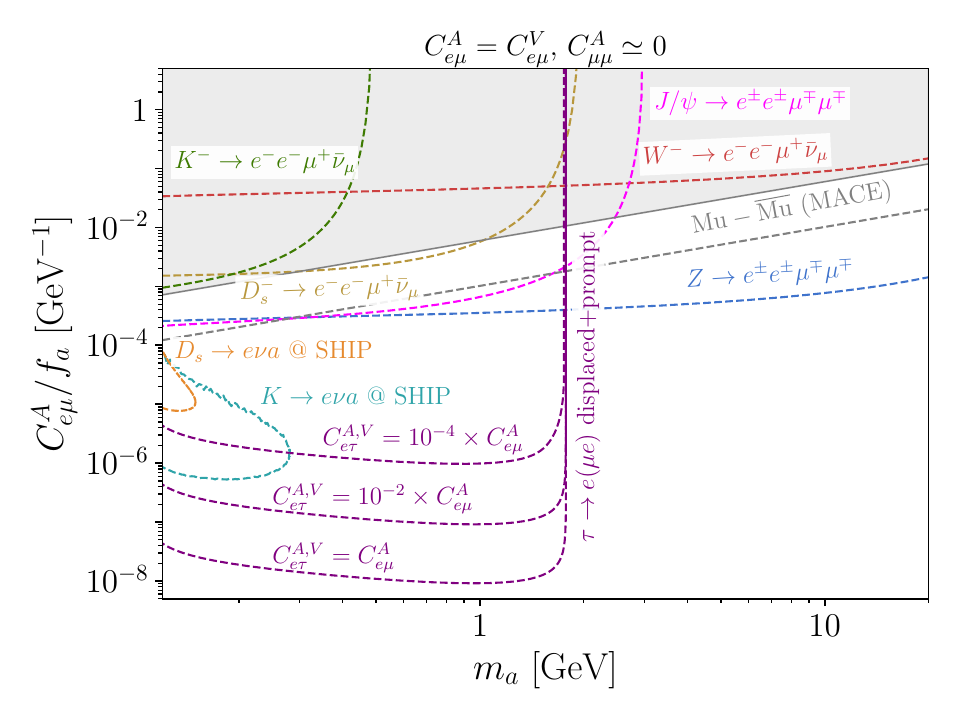}
	\caption{\footnotesize Projected sensitivities and existing bounds in {\bf Scenario A}, assuming a negligible flavour-diagonal muon coupling, $C^A_{\mu\mu}\ll C^{V,A}_{e\mu}$. The upper panel shows the case of predominantly LH ALP couplings, for which $SU(2)_L$-breaking effects can arise; the lower panel shows the corresponding RH case. The grey shaded region indicate the parameter space excluded by current low-energy bounds, dominated by muonium--antimuonium oscillation (see Eq.~(\ref{eq:muonium}) and related discussion), while the dashed coloured curves show the projected sensitivity of the future searches discussed in this work. In this scenario, the ALP decays dominantly through the two LFV modes $a\to e^\pm\mu^\mp$, with ${\rm BR}(a\to e^-\mu^+)={\rm BR}(a\to e^+\mu^-)\simeq 1/2$. The kaon decay sensitivity is from NA62, assuming they can have a reach ${\rm BR}(K^- \to \bar{\nu}_\mu e^-  (a\to e^- \mu^+))\lesssim 10^{-11}$ comparable to other LFV kaon decays~\cite{Bician:2022lsm}. The $Z, D_s$ and $W$ curves show the projected $90\%$ exclusion limits at CEPC/FCC-ee, estimated using Eq.~(\ref{eq:neventsZMW}) with $N_Z=4\times10^{12},\ N_W=2\times 10^8, N_{D_s}=1.8\times 10^{11}$~\cite{Ai:2024nmn}. The $J/\psi$ line is the projected sensitivity of STCF with $N_{J/\psi}=10^{13}$~\cite{Achasov:2023gey}. The $D_s\to e \nu a$ and $K\to e \nu a$ curves are the SHiP limits obtained from Eq.~(\ref{eq:signalSHIP}), more details are given in Appendix \ref{app:beamdump}. The purple dashed lines show the projected limits obtained from the combination of searches for $\tau \to ea\,(\to\mu e)$ with displaced vertices ($1~\mathrm{mm}<l_a<1~\mathrm{m}$), and prompt decays ($l_a<1~\mathrm{mm}$) at Belle~II and Tera-$Z$ factories, for different values of the ratio $C^A_{e\tau}/C^A_{e\mu}$. As discussed at the end of Section~\ref{subsec:colliders} and shown in Fig.~\ref{fig:displaced}, the projected reach of Belle~II is slightly better than the one of FCC-ee/CEPC in the long-lived ALP regime.}
	\label{fig:ScenarioA}
\end{figure}

\begin{figure}[!htp!]
	\centering
	\includegraphics[width=0.64\textwidth, trim=0 0 0.5cm 0.5cm, clip]{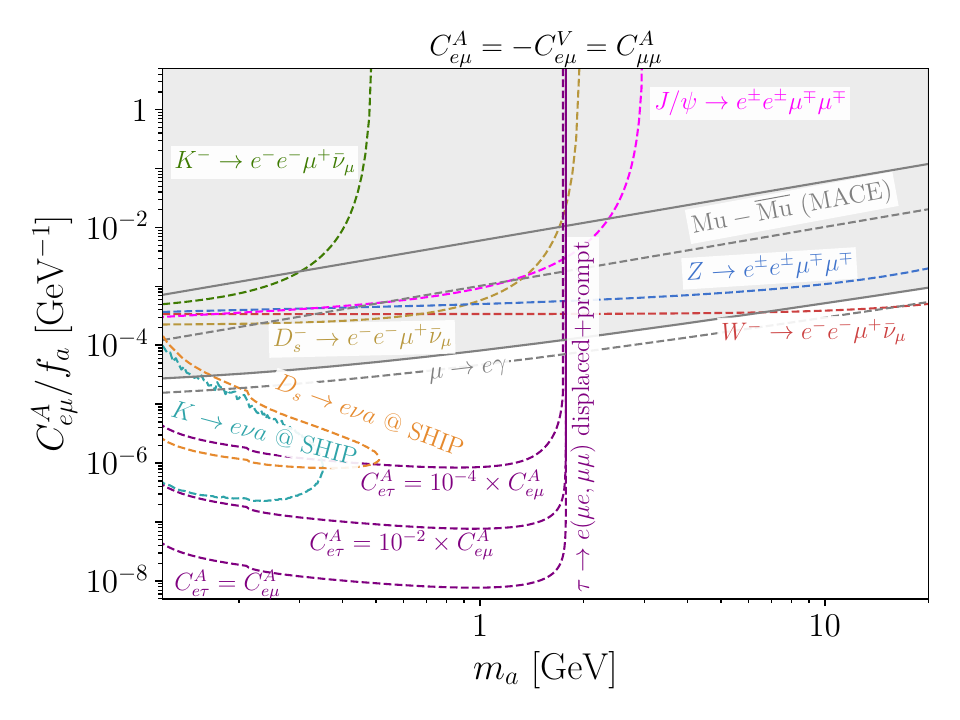}
	\includegraphics[width=0.64\textwidth, trim=0 0 0.5cm 0.5cm, clip]{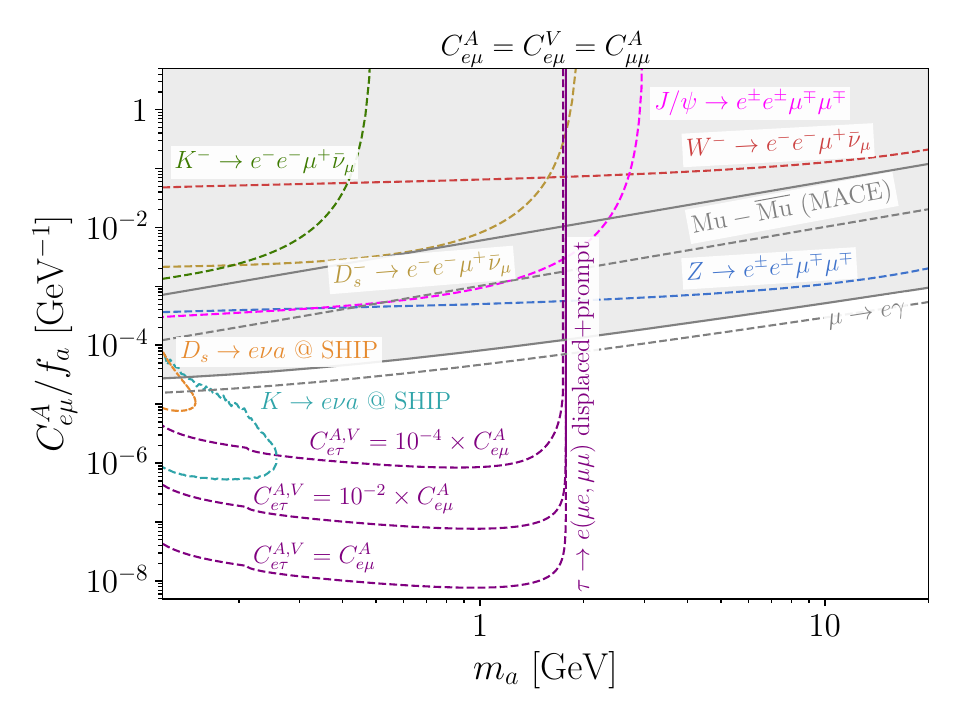}
	\caption{\footnotesize Projected sensitivities and existing bounds in {\bf Scenario B}, assuming a flavour-diagonal muon coupling of the same order as the LFV one, $C^A_{\mu\mu}\simeq C^{V,A}_{e\mu}$. The upper panel corresponds to predominantly LH ALP couplings and the lower panel to predominantly RH ones. In this scenario, strong indirect limits arise from loop-induced $\mu\to e\gamma$ and future $\mu\to eee$ (see Section \ref{ssec:lowenergy}), which exclude a large fraction of the parameter space. The dashed coloured curves show the projected reach of the direct searches proposed in this work (see the caption of Figure \ref{fig:ScenarioA} for more details).  Because the decay channel $a\to\mu^+\mu^-$ is also present with a comparable rate, the branching ratio into each LFV mode is reduced to ${\rm BR}(a\to e^-\mu^+)= {\rm BR}(a\to e^+\mu^-)\simeq 1/4$.
    }
	\label{fig:ScenarioB}
\end{figure}

The corresponding reaches in the ALP parameter space are shown in Figures~\ref{fig:ScenarioA} and~\ref{fig:ScenarioB}, referring respectively to {\bf Scenario A} and {\bf Scenario B}. In both figures, the upper panel corresponds to purely LH couplings and the lower panel corresponds to the RH ones.

The main phenomenological difference between the two scenarios is the impact of the indirect constraints arising from processes induced by virtual ALP exchange. In {\bf Scenario~A}, where the flavour-diagonal muon coupling is negligible, the dominant bound comes from muonium--antimuonium oscillation. In {\bf Scenario~B}, instead, the simultaneous presence of $C^{A}_{\mu\mu}$ and $C^{V,A}_{e\mu}$ induces the radiative decay $\mu\to e\gamma$ at one loop, and correspondingly also strong limits from future $\mu\to eee$ and $\mu\to e$ conversion searches, as discussed in Section~\ref{ssec:lowenergy}. These indirect constraints are stronger than in {\bf Scenario~A} and, as shown in \fref{fig:ScenarioB}, already exclude most of the parameter space that could otherwise be probed by the LFV meson and gauge-boson decays considered here. The main exception is represented by $W$ decays in the LH scenario, where $SU(2)_L$-breaking effects can be sizeable. In this case, the reach can extend beyond the projected sensitivity of future $\mu\to e$ facilities for ALP masses $m_a\gtrsim 10$~GeV.

A further quantitative difference between the two scenarios is the ALP branching ratio into the specific LFV final state selected in our searches. In {\bf Scenario~A}, where no sizeable $\mu\mu$ coupling is present, the visible ALP decays are dominated by the two charge-conjugated LFV modes, so that
\begin{flalign}
	 {\rm\bf Scenario\ A:}\qquad
	{\rm BR}(a \to e^- \mu^+)={\rm BR}(a \to e^+ \mu^-)=\frac{1}{2}\,. &&
\end{flalign}
In {\bf Scenario~B} the additional decay channel $a\to \mu^- \mu^+$ is also open with a comparable rate, so that the branching ratio into each LFV mode is correspondingly reduced:
\begin{flalign}
	{\rm \bf Scenario\ B:}\qquad
	{\rm BR}(a \to e^- \mu^+)={\rm BR}(a \to e^+ \mu^-)\simeq \frac{1}{2}\,{\rm BR}(a \to \mu^- \mu^+)\simeq \frac{1}{4}\,. &&
\end{flalign}
This suppression weakens the sensitivity of direct searches based on the $a\to e\mu$ final state, although the qualitative picture remains unchanged.

Searches at SHiP could probe the complementary region with couplings $C^{A}_{e\mu}/f_a\lesssim 10^{-4}\ \mathrm{GeV}^{-1}$, although the accessible mass range is limited by the relative short ALP decay length. The sensitivity is especially interesting in the LH case, where meson production may be enhanced by the weak-breaking effects. In the RH case, the reach from $D_s$ decays is more limited due to the smaller branching fractions, whereas kaon decays can remain competitive because of the much larger number of kaons produced in the target and the milder $(m_\mu/m_K)^2$ suppression. More details on how we estimate the SHiP sensitivities are given in Appendix \ref{app:beamdump}.

In all benchmark cases analysed here, if non-zero $\ell\tau$ couplings are also present, the combination of prompt and displaced ALP searches at Belle~II and future $e^+e^-$ colliders provides a powerful probe of $e\mu$ interactions in the mass region $m_a<m_\tau$. This remains true even when the $\ell\tau$ couplings are  comparatively suppressed with respect to $e\mu$ ones, as shown in the figures where we display three benchmark lines where the ratio $C^A_{e\tau}/C^A_{e \mu}$ is varied between $1$ and $10^{-4}$. Even in the latter case, these searches can still probe up to $C^{A}_{e\mu}/f_a\gtrsim 10^{-6}\ \mathrm{GeV}^{-1}$.

\section{Conclusions}
\label{conclu}

In this work, we have investigated new strategies to probe ALPs with LFV couplings in the comparatively weakly constrained mass regime $m_a > m_\mu$. In this region, the strong bounds from exotic muon decays disappear, while the opening of the decay channel $a\to e\mu$ implies that the ALP is typically short-lived and gives rise to visible and virtually background-free final states. This motivates a qualitatively different search programme, based on the production of on-shell ALPs in processes involving virtual muons emitted in weak decays of mesons and electroweak gauge bosons, as well as in LFV $\tau$ decays when couplings to the third lepton generation are present.

We have first derived the production rates relevant for this programme. In particular, we studied the three-body decays $M^- \to e^- \bar\nu_\mu a$ and $W^- \to e^- \bar\nu_\mu a$, highlighting the role of the chiral structure of the ALP couplings and of possible departures from the $SU(2)_L$-invariant limit. In the latter case, the effective weak vertex involving the ALP can remove the usual helicity suppression and significantly enhance the rates. We also computed the corresponding production channels in neutral boson decays, such as $Z\to e^\pm\mu^\mp a$, as well as in heavy-quarkonium decays, and discussed the complementary role of two-body LFV $\tau$ decays, $\tau\to \ell a$, in the mass window $m_\mu \lesssim m_a \lesssim m_\tau$.

At the phenomenological level, we have shown that these production mechanisms lead to striking signatures. For meson and $W$ decays, the signal consists of two same-sign electrons, an oppositely charged muon, and missing energy, e.g.~$W^- \to \bar \nu_\mu   e^-  e^- \mu^+$; for $Z$ and quarkonium decays, one obtains four charged leptons with a resonant $e\mu$ pair, such as $Z \to e^+\mu^- e^+\mu^-$. Such final states violate the conservation of the lepton family numbers in a very distinctive way (even considering that the neutrino flavour cannot be observed) and can therefore be searched for in an essentially background-free regime. We have used this feature to assess the projected sensitivity of future $e^+e^-$ colliders, Belle~II, and proton fixed-target or beam-dump experiments. The resulting sensitivity reach significantly depends on the ALP coupling pattern. In scenarios with negligible flavour-conserving couplings to muons, the dominant indirect constraint is provided by searches for muonium--antimuonium conversion, leaving substantial room for direct ALP searches. By contrast, if flavour-conserving and flavour-violating muon couplings are both present with comparable size, loop-induced $\mu\to e\gamma$ and future $\mu\to eee$ searches strongly reduce the available parameter space.

Our results show that future high-luminosity $e^+e^-$ colliders have an excellent discovery potential for LFV ALPs in this mass region. In particular, Tera-$Z$ factories such as CEPC and FCC-ee will provide access to very large samples of $Z$ bosons, $\tau$ leptons and $D_s$ mesons, while the clean environment of a lepton collider is well suited to reconstruct the characteristic LFV final states considered here. In addition, fixed-target facilities such as NA62 and beam-dump experiments such as SHiP can probe complementary regions of parameter space, especially in scenarios where the ALP is produced in kaon or $D_s$ meson decays and is sufficiently long-lived to decay inside the detector. If LFV couplings to the $\tau$ lepton are present, Belle~II and future $Z$ factories are also sensitive through both prompt and displaced ALP decays, providing indirect access to the underlying $e\mu$ interactions.

Overall, our analysis shows that LFV ALP production in meson, gauge-boson, quarkonium and $\tau$ decays opens up a broad and largely unexplored avenue to test ALP interactions with charged leptons above the muon threshold. Our main results are illustrated by Figures~\ref{fig:ScenarioA} and~\ref{fig:ScenarioB}, which show that, across different and complementary ranges of the ALP mass, CEPC/FCC-ee can test energy scales up to $f_a \approx 5 \TeV$ through meson and gauge boson decays, with a similar sensitivity expected at STCF from searches for LFV $J/\psi$ decays;
SHiP can reach $f_a \approx 10^6 \GeV$, while, in presence of LFV $\tau$ interactions, Belle~II and CEPC/FCC-ee can be sensitive to scales up to $f_a \approx 10^8 \GeV$. Hence, the combination of current low-energy bounds with the direct probes proposed here allows one to cover significant portions of the viable parameter space and, in particular, to access scenarios that are not constrained by conventional exotic-muon searches. We therefore conclude that LFV final states involving an intermediate ALP deserve to be included among the benchmark targets of future flavour and intensity-frontier experiments.

\acknowledgments
The Feynman diagrams shown in this work were created using Tikz-Feynman~\cite{Ellis:2016jkw}.
We would like to thank Admir Greljo and Mauro Raggi for insightful discussions. 
MA, LC and MF acknowledge the INFN Laboratori Nazionali di Frascati for hospitality and partial financial support during the completion of this project.
MA acknowledges financial support from the Spanish Grant PID2023-151418NB-I00 funded by
MCIU/AEI/10.13039/ 501100011033/ FEDER, UE and from Generalitat Valenciana
projects CIPROM/2021/054 and CIPROM/2022/66.
MF acknowledges financial support from  the Cluster of Excellence \textit{PRISMA}$^{++}$ (EXC 2118/2, Project ID 390831469).
FM acknowledges financial support from the European Union-Next Generation EU and by the Italian Ministry of University and Research (MUR) via the PRIN 2022 project No.~2022K4B58X-AxionOrigins.

\section*{Appendix}

\appendix

\section{Details on weak violation model building}
\label{app:weak}

In this appendix, we comment on the possible origin of the weak-breaking effects introduced in Ref.~\cite{Altmannshofer:2022ckw} and discussed above, which can give rise to a misalignment between the ALP couplings to LH charged leptons and neutrinos, such that
\begin{equation}
	C^V_{ij}-C^A_{ij}-C^\nu_{ij}\neq 0 \, .
\end{equation}
At the effective level, such a misalignment can arise in any model that generates higher-dimensional operators of the form
\begin{equation}
	\Delta\mathcal L_{\rm eff}
	=
	\frac{\partial_\alpha a}{f_a}\,
	\frac{\kappa_{ij}}{\Lambda^2}\,
	(\bar L_i H)\gamma^\alpha(H^\dagger L_j)
	+{\rm h.c.}\, ,
	\label{eq:aeHLeff}
\end{equation}
where $L_i$ denotes the lepton doublet, $L_i=(\nu_{Li}, \ell_{Li})^T$ .
After electroweak symmetry breaking, the Higgs field $H$ takes the form $H=(0,(v+h)/\sqrt2)^T$ and this operator matches onto
\begin{equation}
	\Delta\mathcal L_{\rm eff}
	=
	\frac{\partial_\alpha a}{2f_a}\,
	\frac{v^2}{\Lambda^2}\,
	\kappa_{ij}\,
	\bar \ell_{Li}\gamma^\alpha \ell_{Lj}
	+\cdots \, .
	\label{eq:aeHLeffEWSB}
\end{equation}
We therefore observe that ALP couplings to LH charged leptons are generated, while couplings to neutrinos are absent. One thus obtains
\begin{equation}
	\Delta C^\nu_{ij}=0\,,
	\qquad
	\Delta C^V_{ij}=\frac{v^2}{2\Lambda^2}\kappa_{ij}\,,
	\qquad
	\Delta C^A_{ij}=-\frac{v^2}{2\Lambda^2}\kappa_{ij}\,,
	\label{eq:matchingweakbreaking}
\end{equation}
so that Eq.~(\ref{eq:aeHLeff}) directly generates the combination controlling the effective $W$ vertex in Eq.~(\ref{eq:Wvertex}).

\paragraph{A minimal UV completion.}
A simple UV completion can be obtained by introducing a heavy vector-like lepton transforming as $E_{L,R}\sim(1,1,-1)$ under $SU(3)\times SU(2)_L\times U(1)_Y$, and charged under the PQ symmetry. The relevant interactions are
\begin{equation}
	\mathcal L_{\rm UV}\supset
	\left(
	m_E\,\bar E_L E_R
	+y_E^i\,\bar{L}_i H E_R
	+{\rm h.c.}
	\right)
	+\frac{\partial_\alpha a}{2f_a}\,c_E\,\bar E_L\gamma^\alpha E_L \, .
	\label{eq:UVVLL}
\end{equation}
When the Higgs acquires a vacuum expectation value, the Yukawa interactions mix the SM LH charged leptons with the LH component of the heavy state, with mixing angles
\begin{equation}
	\theta_i \simeq \frac{y_E^i\,v}{\sqrt2\,m_E}\, .
\end{equation}
Since the ALP couples to $E_L$, this mixing induces a current for the light charged leptons,
\begin{equation}
	\Delta\mathcal L
	=
	\frac{\partial_\alpha a}{2f_a}\,
	c_E\,\theta_i\theta_j^*\,
	\bar \ell_{Li}\gamma^\alpha \ell_{Lj}
	=
	\frac{\partial_\alpha a}{2f_a}\,
	c_E\,
	\frac{v^2\,y_E^i y_E^{j*}}{2m_E^2}\,
	\bar \ell_{Li}\gamma^\alpha \ell_{Lj}\, .
	\label{eq:mixingresult}
\end{equation}
Equivalently, integrating out the heavy lepton $E$ at tree level generates Eq.~(\ref{eq:aeHLeff}) with
\begin{equation}
	\frac{\kappa_{ij}}{\Lambda^2}=\frac{c_E\,y_E^i y_E^{j*}}{2 m^2_E}\,,
	\label{eq:kappamatching}
\end{equation}
which gives
\begin{equation} \label{eq:weakbreakingALPc}
	\Delta(C^V_{ij}-C^A_{ij}-C^\nu_{ij})
	=
	c_E\,\frac{v^2\,y_E^i y_E^{j*}}{2m_E^2}\, .
\end{equation}
Therefore, obtaining order-one weak-breaking effects requires sizeable mixing between the SM leptons and the vector-like lepton, together with a not-too-large vector-like lepton mass. While sub-TeV singlet leptons mixing with either electrons or muons are experimentally allowed, see Ref.~\cite{ATLAS:2024mrr}, the same Yukawa couplings $y_E^i$ also generate lepton dipole operators at one loop. In the Standard Model Effective Theory (SMEFT), these operators are
\begin{equation}
	\mathcal{O}_{eB}^{ij}=(\bar L_i \sigma^{\mu\nu} e_j)\,H\,B_{\mu\nu}\,,
	\qquad
	\mathcal{O}_{eW}^{ij}=(\bar L_i \sigma^{\mu\nu} e_j)\,\tau^I H\,W^I_{\mu\nu}\,,
\end{equation}
and for a charged-singlet vector-like lepton their coefficients result~\cite{OlgosoRuiz:2023npk}
\begin{equation}
	[C_{eB}]_{ij}
	=
	-\frac{g^\prime}{16\pi^2}\,
	\frac{1}{12m_E^2}\,
	y_E^i y_E^{k*}(Y_e)_{kj}\,,
	\qquad
	[C_{eW}]_{ij}
	=
	-\frac{g}{16\pi^2}\,
	\frac{1}{24m_E^2}\,
	y_E^i y_E^{k*}(Y_e)_{kj}\, ,
	\label{eq:dipolematching}
\end{equation}
where $Y_e$ is the SM charged-lepton Yukawa matrix. After electroweak symmetry breaking, these operators induce the photon dipole
\begin{equation}
	\mathcal L \supset
	\frac{v}{\sqrt2}\,[C_{e\gamma}]_{ij}\,
	\bar \ell_{Li}\sigma^{\mu\nu} \ell_{Rj} F_{\mu\nu}
	+{\rm h.c.}\,,
	\qquad \text{with}
    \qquad
	[C_{e\gamma}]_{ij}=c_W[C_{eB}]_{ij}-s_W[C_{eW}]_{ij}\,,
\end{equation}
whose coefficient reads
\begin{equation}
	[C_{e\gamma}]_{ij}
	=
	-\frac{e}{384\pi^2 m_E^2}\,
	y_E^i y_E^{k*}(Y_e)_{kj}\, .
	\label{eq:photonDipole}
\end{equation}
The induced dipole operator is therefore proportional to the same flavour spurion $y_E^i y_E^{j*}$ that controls the weak-breaking ALP coupling in Eq.~\eqref{eq:weakbreakingALPc}. As a result, the current bound \mbox{${\rm BR}(\mu\to e\gamma)<1.5 \times 10^{-13}$}~\cite{MEGII:2025gzr} strongly constrains the size of the off-diagonal $e\mu$ ALP coefficient:
\begin{equation}
	\frac{v^2\,y_E^e y_E^{\mu*}}{2m_E^2}\lesssim  10^{-4}\,.
	\label{eq:boundweakbreakingUV}
\end{equation}
This makes it difficult to obtain large weak-breaking effects in the $e\mu$ sector. 

One possible way to relax this bound is to couple the heavy lepton to electrons, taking
$y_E^\mu=0$ and $y_E^e\neq0$. In this way, before diagonalising the light charged-lepton mass matrix, both the dipole operator and the induced ALP current are flavour diagonal. In particular, the latter reads
\begin{equation}
	\Delta\mathcal L
	=
	\frac{\partial_\alpha a}{2f_a}\,
	c_E\,
	\frac{v^2 |y_E^e|^2}{2m_E^2}\,
	\bar e_L\gamma^\alpha e_L\, .
\end{equation}
If the charged-lepton mass matrix is then diagonalised by a non-trivial LH rotation,
$\ell_L=V_\ell\,\ell_L^\prime$, this current becomes
\begin{equation}
	\Delta\mathcal L
	=
	\frac{\partial_\alpha a}{2f_a}\,
	c_E\,
	\frac{v^2 |y_E^e|^2}{2m_E^2}\,
	(V_\ell^\dagger P_e V_\ell)_{ij}\,
	\bar\ell^\prime_{Li}\gamma^\alpha \ell^\prime_{Lj}\,,
	\label{eq:rotatedcurrent}
\end{equation}
where $P_e={\rm diag}(1,0,0)$. 
The same flavour rotation also misaligns the diagonal electron dipole operator $[C_{e\gamma}]_{ee}$ defined in Eq.~\eqref{eq:photonDipole}, thereby generating the $\mu\to e\gamma$ transition. In this case, however, this amplitude is suppressed by the electron mass, so that the bound on the $e\mu$ ALP couplings in Eq.~(\ref{eq:boundweakbreakingUV}) can be relaxed by a factor $(m_\mu/m_e)\simeq 200$. This alternative alleviates, while not removing, the general tension: namely, in simple UV completions, the weak-breaking LFV ALP coupling can also be constrained by the contributions of the heavy states to LFV observables. 

More complicated UV completions might allow for larger weak-breaking effects at the price of tuning (or the introduction of additional symmetries). For instance, if three generations of vector-like singlets $E$ are introduced, the couplings $y_E$ in Eq.~\eqref{eq:UVVLL} become a $3\times 3$ matrix. 
{If the matrix entering in Eq.~\eqref{eq:photonDipole}, $y_E M^{-2}_E y_E^\dagger Y_e$, can be diagonalised in the same basis where $Y_e$ is diagonal (which requires a very specific flavour structure, either accidental or enforced by some underlying symmetry), the flavour changing dipole entries vanish, removing the stringent $\mu\to e\gamma$ constraints. Note that the ALP couplings are instead parametrized by the combination $y_E c_E M_E^{-2}  y^\dagger_E$, which can be off-diagonal if the $c_E$ PQ charges are not universal. }

\section{Simulation of the SHiP sensitivity}
\label{app:beamdump}
In this appendix, we provide further details on the simulation used to estimate the ALP flux from the decay $M\to e \bar{\nu}_\mu a$, where the meson $M=D_s,K$ is produced in $pp(pn)$ collisions on the SHiP target, and the ALP is subsequently observed via its decays inside the downstream decay volume.
We recall that, in order to be detected, the ALP must first travel a distance $L$ from its production point to the entrance of the fiducial volume and subsequently decay within it. The corresponding probability is
\begin{equation}
	P_a = \exp\left(-\frac{\Gamma_a L}{\gamma_a \beta_a}\right)
	\left[
	1-\exp\left(-\frac{\Gamma_a \lambda}{\gamma_a \beta_a}\right)
	\right]\,,\label{eq:probabilityALPSHIP}
\end{equation}
where $\lambda$ is the path length covered inside the fiducial volume, $\Gamma_a$ is the ALP decay width, and $\gamma_a=E_a/m_a$ is the boost factor in the lab frame. $L$ and $\lambda$ depend on the direction of flight of the ALP (and hence on the ALP momentum when produced) and crucially on the geometry of the experiment: the fiducial volume begins about $34$ m downstream of the target and is $\approx 50~\mathrm{m}$ long with the shape of a pyramidal frustum, while the upstream and downstream transverse dimensions are approximately $1.0\times2.7~\mathrm{m}^2$ and $4\times6~\mathrm{m}^2$, respectively. The quantities $L$ and $\lambda$ are determined knowing the ALP trajectory, considering the length covered before and after entering the fiducial volume. As a result, the geometrical acceptance of the experimental setup is already encoded in Eq.~(\ref{eq:probabilityALPSHIP}).\footnote{For example, if an ALP has sufficiently large transverse momentum to miss the decay volume, then $\lambda=0$ and therefore $P_a=0$.}

In the rest frame of the decaying meson, the ALP energy distribution to leading order in $m_\mu^2$ is
\begin{align}
	\frac{d\Gamma_{M\to e \nu a}}{dE^*_a}
	=
	\frac{f_{M}^2 G_F^2 m_M|V_{ij}|^2}{48\pi^3 f_a^2}
	\Bigg\{
	|C_L|^2 k_a^{*3}
	&+ \frac{3 m_\mu^2}{2 m_M^2}
	\Bigg[
	k^*_a 
	\left(
	2 |C_L|^2 E^*_a m_M + |C_R|^2 s_{e\nu}
	\right)\nonumber\\
	&+ |C_L|^2 m_a^2 m_M 
	\log\left(
	\frac{E^*_a - k^*_a}
	{E^*_a + k^*_a}
	\right)
	\Bigg]
	\Bigg\}\,,
	\label{eq:energyalp}
\end{align}
where
\begin{equation}
	k^*_a = \sqrt{E_a^{*2} - m_a^2}\,,
	\qquad
	s_{e\nu} = m_M^2 - 2E_a^*m_M + m_a^2 \,.
\end{equation}
The integral of Eq.~(\ref{eq:energyalp}) over the ALP energy in the $M$ rest frame reproduces Eq.~(\ref{eq:mesonrate}).

\begin{figure}[t]
	\centering
	\includegraphics[width=0.7\textwidth]{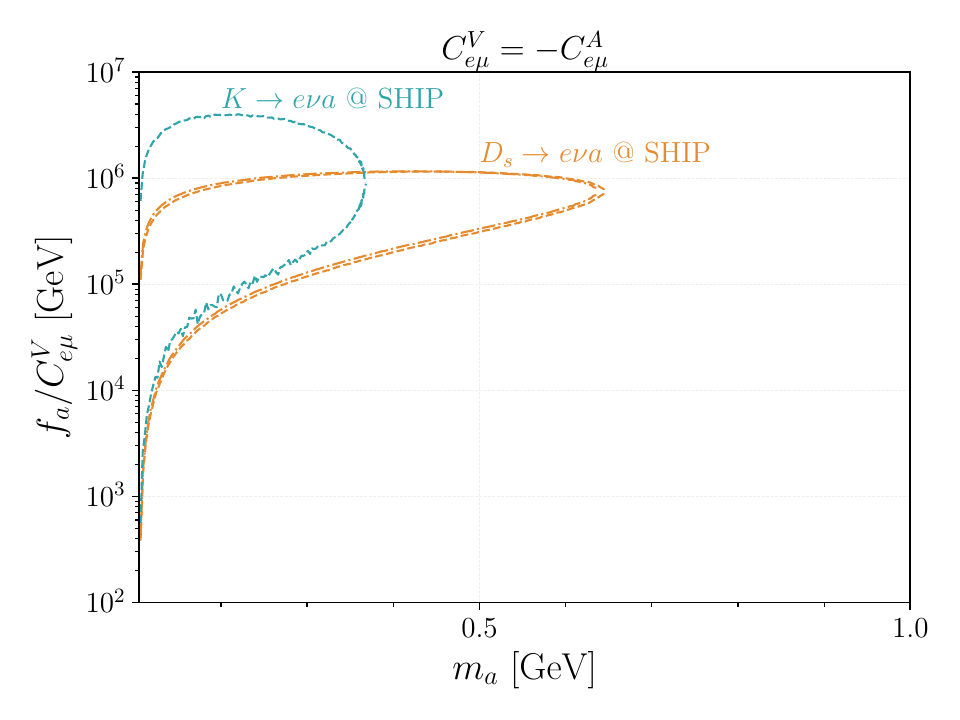}
	\caption{Expected SHiP sensitivity in the case of ALP left-handed $e\mu$ couplings, for ALP production via the decays $D_s \to e \nu_\mu a$ and $K \to e \nu_\mu a$. The two $D_s$ curves correspond to the limits obtained using the distributions in Eq.~(\ref{eq:DistributionEmaetal}) (dot-dashed) and Eq.~(\ref{eq:DistributionSimplified}) (dashed). This figure shows a zoomed-in view of the SHiP sensitivities presented in the upper panel of Fig.~\ref{fig:ScenarioA}, plotting the inverse in the $y$-axis.}
	\label{fig:zoombeamdumpLH}
\end{figure}

\paragraph{$D_s$ mesons.}
 For a high-energy proton beam with $E_b \gg m_p$ impinging on a fixed target, the centre-of-mass (COM) energy $\sqrt{s}$ of the underlying $pp$ collision is approximately
\begin{equation}
	\sqrt{s} \simeq \sqrt{2 E_b m_p}\,,
\end{equation}
while the velocity of the COM frame with respect to the laboratory frame, taking the $z$ axis along the beam direction, is
\begin{equation}
	\beta^{\mathrm{CM}}_{z} \simeq 1 - \frac{m_p}{E_b}\,.
\end{equation}
In the COM frame of the $pp$ collision, the kinematic distribution of the produced $D_s$ mesons can be parametrized using the empirical form~\cite{Lourenco:2006vw}
\begin{equation}
	\frac{dN_{D_s}}{dp_T\,dx_F}
	\propto
	(1 - |x_F|)^n \exp(-b p_T^2)\,,
	\label{eq:DistributionSimplified}
\end{equation}
where $x_F$ is the fraction of the longitudinal momentum carried by the $D_s$ meson in the collision COM frame, defined as $x_F = p_z / p_z^{\max}$, while $p_T$ is the transverse momentum. The parameters $n$ and $b$ are fitted to experimental data and depend on the beam energy. For SPS (400 GeV proton beam), we use $n=5.8$ and $b=0.96\ {\rm GeV}^{-2}$~\cite{Schubert:2024hpm}.
\begin{figure}[t]
	\centering
	\includegraphics[width=0.7\textwidth]{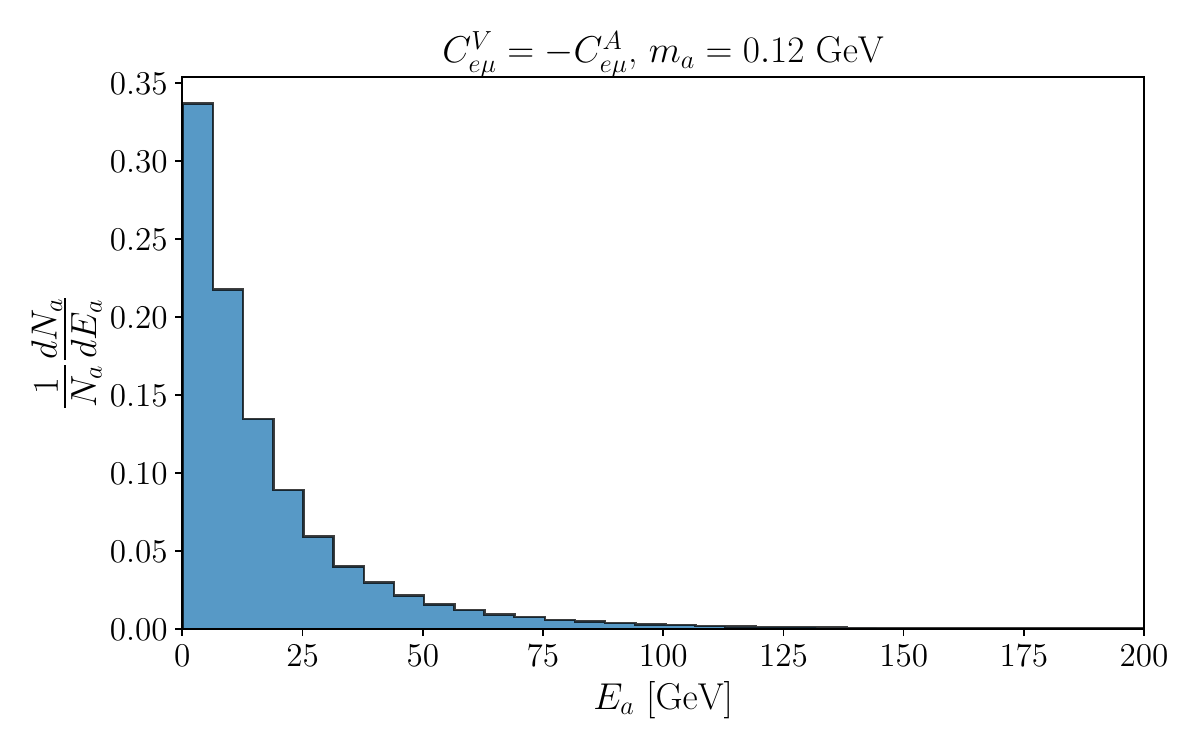}
	\caption{Energy distribution in the laboratory frame of ALPs produced in the LFV decay $D_s\to e \nu_\mu a$, for $m_a=0.12$ GeV and left-handed ALP couplings. The result is based on $N_a=10^5$ simulated events.}
	\label{fig:alpenergy}
\end{figure}
We sample the $D_s$ kinematics from the distribution in Eq.~(\ref{eq:DistributionSimplified}) and the ALP energy in the $D_s$ rest frame from Eq.~(\ref{eq:energyalp}). If the ALP is produced in the $D_s$ frame with a four-momentum
\begin{equation}
	p^*_a=
	\left(
	E_a^*,
	k_a^*\sin\theta_a^*\cos\phi_a^*,
	k_a^*\sin\theta_a^*\sin\phi_a^*,
	k_a^*\cos\theta_a^*
	\right)\,,
\end{equation}
where the angles $\phi_a^*$ and $\theta_a^*$ are sampled uniformly respectively in the intervals $[0,2\pi]$ and $[0,\pi]$,\footnote{Since the parent meson is a pseudoscalar, the ALP is emitted isotropically in the meson rest frame.} then the ALP four-momentum in the laboratory frame is obtained by performing the inverse boosts
\begin{equation}
	p_a=\Lambda (-\beta^{\rm CM}_z)\Lambda(-\beta_{D_s}(p_T, x_F))p^*_a\,,
\end{equation}
where $\beta_{D_s}$ is the $D_s$ velocity in the COM of the $pp$ collisions. For each value of $m_a$ analysed, we generate $N_a=10^5$ ALP events following the procedure described above (an example of the ALP energy distribution we obtain is given in Figure \ref{fig:alpenergy}), and evaluate numerically the ALP signal-event integral in Eq.~(\ref{eq:signalSHIP}) summing over the Monte Carlo sample
\begin{equation}
	N_{\rm sig}\simeq \frac{1}{N_a}\sum_{\rm sample}
	N_{D_s}\times {\rm Br}(D_s\to e \nu_\mu a)\times P_a\,, \label{eq:montecarloint}
\end{equation}
where $P_a$ is the probability of the ALP decaying within the SHiP decay volume, defined in Eq.~\eqref{eq:probabilityALPSHIP}. The total number of produced $D_s$ mesons is
\begin{equation}
	N_{D_s}=N_{\rm POT}\frac{\sigma_{c\bar{c}}}{\sigma^{\rm inel}_{pp}} f_{c\to D_s}\,,
\end{equation}
where $N_{\rm POT}=6\times 10^{20}$ is the expected number of protons on target at SHiP, $\sigma_{c\bar{c}}$ is the charm-pair production cross section, $\sigma^{\rm inel}_{pp}$ is the total inelastic $pp$ cross section, and $f_{c\to D_s}$ is the fragmentation fraction of the $cc$ pairs into $D_s$. Using $\sigma_{c\bar{c}}/\sigma^{\rm inel}_{pp}\simeq 7.2\times 10^{-4}$~\cite{Schubert:2024hpm} and $f_{c\to D_s}\simeq 0.079$ \cite{Gladilin:2014tba}, we obtain
\begin{equation}
	N_{D_s}\simeq N_{\rm POT}\times (5\times 10^{-5})\,.\label{eq:NDs}
\end{equation}
Other simulations quote a number of $D_s$ produced at SHiP that can be an order of magnitude larger \cite{Ovchynnikov:2023cry}; here, however, we adopt the more conservative estimate of Eq.~(\ref{eq:NDs}). This normalization is also compatible with the $D_s$ yield obtained in the simulation of Ref.~\cite{Ema:2025bww}, although they find that the $D_s$ kinematical distributions are better described by the alternative parametrization
\begin{equation}
	\frac{dN_{D_s}}{dx_F\,dp_T}
	\propto
	\left[
	\sum_{i=1}^{3} g_i \Gamma(p_T^{m_i},\alpha_i;\lambda_i)
	\right]
	\left[
	\sum_{j=1}^{3} c_j \exp\left(-a_j |x_F|^{n_j}\right)
	\right]\,,
	\label{eq:DistributionEmaetal}
\end{equation}
where $\Gamma(x,\alpha,\lambda)=
	\lambda^{\alpha}x^{\alpha-1}e^{-\lambda x}/\Gamma(\alpha).$ We also simulate the $D_s$ spectrum using Eq.~(\ref{eq:DistributionEmaetal}) and the best-fit parameters reported in Table~II of Appendix~C of Ref.~\cite{Ema:2025bww}, and find similar results for the exclusion limit in the $m_a\ {\rm vs}\ C^A_{e\mu}/f_a$ plane (see for instance Figure \ref{fig:zoombeamdumpLH}, where the $D_s\to e \nu_\mu a$ lines obtained sampling the distributions in Eq.~(\ref{eq:DistributionEmaetal}) and Eq.~(\ref{eq:DistributionSimplified}) are shown together).

\paragraph{Kaons.} 
If $D_s$ mesons are sufficiently short-lived to decay promptly after being produced in $pp$ collisions, kaons, being significantly longer-lived, can instead undergo multiple scatterings within the SHiP target. As a consequence, most kaons are either absorbed or brought to rest inside the target, while those that escape are subsequently stopped in the hadron absorber located downstream of the target. Nevertheless, a fraction of kaons may decay into ALPs via the $e\mu$ coupling ($K \to e \nu_\mu a$) before being absorbed. In such cases, the produced ALP may subsequently be observed through its decay within the SHiP fiducial volume. As discussed in Section \ref{subsec:beamdump}, the kaons relevant for this analysis are those that retain a sufficiently large forward momentum, $p_z$, such that the emitted ALP is directed toward the downstream decay volume. By contrast, kaons decaying only after coming to rest are expected to give a negligible contribution, since the probability for the ALP to be emitted in the appropriate direction is small. To estimate the number of such “useful” kaons, and to determine their kinematic distributions, we simulate the interaction of the SPS proton beam with a SHiP-like target in GEANT4~\cite{GEANT4:2002zbu}, considering the target configuration described in Ref.~\cite{SHiP:2015vad}, consisting of a hybrid molybdenum and tungsten target of total length $\simeq 1.2\  \mathrm{m}$, corresponding to roughly $\simeq 10$ nuclear interaction lengths. Our analysis follows closely that of Ref.~\cite{Gorbunov:2020rjx}.

We simulate $N_{pp}=3\times10^5$ proton–target collisions and find an average yield of $9.87$ kaons per POT, of which $6.97$ are $K^+$ and $2.90$ are $K^-$. Among these, we find that approximately $0.03$ kaons per POT decay in flight with laboratory-frame forward momentum $p_z > 1\,\mathrm{GeV}$, whereas the remainder are either absorbed in the target or arrive at the absorber ($7.02$ kaons per POT) or decay in the target only after being stopped ($2.82$ kaons per POT). For the resulting sample of $10^4$ kaons decaying in flight, we record their momentum distribution in the laboratory frame. We then force their decay into ALPs, sampling the ALP energy distribution in the kaon rest frame according to Eq.~(\ref{eq:energyalp}). Considering isotropic ALP emission in the COM frame, we subsequently boost the ALP momentum back to the laboratory frame in order to obtain the corresponding ALP distribution.
The number of signal events is then calculated, similarly to Eq.~(\ref{eq:montecarloint}), as
\begin{equation}
	N_{\rm sig}\simeq \frac{1}{N_a}\sum_{\rm sample}
	N_{K}\times {\rm Br}(K\to e \nu_\mu a)\times P_a\,,
\end{equation}
where the number of kaons is given by
\begin{equation}
	N_K \simeq N_{\rm POT}\times 0.03\,,
\end{equation}
with $0.03$ denoting the average number of kaons per proton on target that decay in flight, as obtained from our simulation. The expected $90\%$ exclusion limit, corresponding to $N_{\rm sig}\leq 2.3$, is shown in Figure \ref{fig:zoombeamdumpLH} for the case of LH $e \mu$ ALP couplings.  

\bibliographystyle{JHEP}
\bibliography{References}

\end{document}